\documentclass[iop]{emulateapj}

\usepackage{color}
\usepackage{float}
\usepackage{epstopdf}
\usepackage{graphicx}
\usepackage[utf8]{inputenc}
\usepackage{natbib}
\renewcommand{\arraystretch}{1.5}
\shortauthors{Martis et al.}
\shorttitle{Physical Properties of Massive, Dusty Galaxies}
\shortauthors{Martis et al.}

\begin{document}

\title{Stellar and Dust Properties of a Complete Sample of Massive Dusty Galaxies at $1 \le z \le 4$ from MAGPHYS Modeling of UltraVISTA DR3 and Herschel Photometry}


\author{Nicholas S. Martis\altaffilmark{1}, Danilo M. Marchesini\altaffilmark{1}, Adam Muzzin\altaffilmark{2}, Mauro Stefanon\altaffilmark{3}, Gabriel Brammer\altaffilmark{4}, Elisabete da Cunha\altaffilmark{5}, Anna Sajina\altaffilmark{1}, Ivo Labbe'\altaffilmark{6}}

\affil{\altaffilmark{1}Physics and Astronomy Department, Tufts University, Medford, MA 02155}
\affil{\altaffilmark{2}Department of Physics and Astronomy, York University, 4700 Keele St., Toronto, Ontario, Canada, MJ3 1P3} %Adam
\affil{\altaffilmark{3}Leiden Observatory, Leiden University, PO Box 9513, NL-2300 RA, Leiden, The Netherlands} %mauro
\affil{\altaffilmark{4}Space Telescope Science Institute, 3700 San Martin Drive, Baltimore, MD 21218, USA}
\affil{\altaffilmark{5}The Australian National University, Mt Stromlo Observatory, Cotter Rd, Weston Creek, ACT 2611, Australia} %elisabete
\affil{\altaffilmark{6}Centre for Astrophysics and SuperComputing, Swinburne, University of Technology, Hawthorn, Victoria, 3122, Australia} %ivo

\email{nicholas.martis@tufts.edu}

%\altaffiltext{1}{Visiting Astronomer, Cerro Tololo }

\begin{abstract}

We investigate the stellar and dust properties of massive (log$(M_*/M_\odot) \ge 10.5$) and dusty ($A_V \ge 1$) galaxies at $1 \le z \le 4$ by modeling their spectral energy distributions (SEDs) obtained from the combination of UltraVISTA DR3 photometry and \textit{Herschel} PACS-SPIRE data using MAGPHYS. Although the rest-frame U-V vs V-J (UVJ) diagram traces well the star-formation rates (SFR) and dust obscuration (A$_V$) out to $z \sim 3$, $\sim$15-20\% of the sample surprisingly resides in the quiescent region of the UVJ diagram, while $\sim50$\% at $3<z<4$ fall in the unobscured star-forming region. The median SED of massive dusty galaxies exhibits weaker MIR and UV emission, and redder UV slopes with increasing cosmic time. The IR emission for our sample has a significant contribution ($>20\%$) from dust heated by evolved stellar populations rather than star formation, demonstrating the need for panchromatic SED modeling. The local relation between dust mass and SFR is followed only by a sub-sample with cooler dust temperatures, while warmer objects have reduced dust masses at a given SFR. Most star-forming galaxies in our sample do not follow local IRX-$\beta$ relations, though IRX does strongly correlate with A$_V$. Our sample follows local relations, albeit with large scatter, between ISM diagnostics and sSFR. We show that FIR-detected sources represent the extreme of a continuous population of dusty galaxies rather than a fundamentally different population. Finally, using commonly adopted relations to derive SFRs from the combination of the rest-frame UV and the observed 24$\micron$ is found to overestimate the SFR by a factor of 3-5 for the galaxies in our sample.
\end{abstract}


\section{Introduction}
The advent of both wide and deep surveys in the near-infrared (NIR) in recent years has allowed astronomers to trace the buildup of stellar mass over a large fraction of the history of the universe \citep{marchesini09, ilbert13, muzzin13b, tomczak14, davidzon17}. One of the limitations of these surveys, though, lies in the uncertain qualities of dust extinction at high redshift. The amount of light obscured by dust increases with decreasing wavelength. Since NIR surveys select sources at progressively shorter wavelengths with increasing redshift, the importance of corrections to account for dust obscuration grows increasingly important as our understanding grows increasingly incomplete. This issue becomes particularly pronounced when determining the star formation rates (SFRs) of galaxies using the rest frame ultraviolet (UV) and can lead to systematic uncertainties of factors of several \citep{bell02, hao11}.

A complementary view of the star formation history of the universe has been made possible through the development of far-infrared (FIR) telescopes, especially \textit{Herschel} \citep[see][for a review]{lutz14}. IR surveys can directly trace the thermal emission of dust grains, allowing a measurement of the amount of obscured star formation. It is now generally regarded as best practice to include both UV and IR measurements to robustly determine the SFR of a galaxy when possible.

Along with the development of FIR telescopes came the discovery of a class of galaxies named for their detection at sub-millimeter wavelengths (SMGs) \citep{smail97, barger98, blain02}. These galaxies are massive, highly star-forming, and heavily obscured. The poor spatial resolution of FIR surveys combined with faint or undetected counterparts at shorter wavelengths for many SMGs makes a systematic study of their panchromatic spectral energy distributions (SEDs) challenging. Compounding these issues is the limited depth of FIR surveys that sample a statistical number of galaxies at high redshift, which restricts these studies to luminous, vigorously star-forming sources.  \citet{schreiber15} estimate that individual \textit{Herschel} detections account for less than 50\% of the SFR density above $z=2$ for their sample in COSMOS which has been observed to moderate depths, so linking these sources to the overall picture of the star formation history of the universe remains challenging. 

In the local universe, the most luminous infrared galaxies are the results of gas-rich major mergers. These systems tend to be massive and very dusty. In the high redshift universe where galaxies are both more compact and gas rich \citep[e.g.][]{tacconi13, vanderwel14, genzel15}, the importance of mergers in generating the observed high SFRs in SMGs is still unclear \citep{hayward11, hayward12, magnelli12}. Given their already substantial stellar masses and high SFRs, these galaxies have the potential to become very massive in the local universe as they evolve \citep{gonzalez11}. Indeed \citet{marchesini14} showed that the typical progenitors of today's most massive galaxies are massive, dusty star-forming galaxies. Identifying the properties of these progenitors will thus provide important constraints for models of galaxy evolution.

This paper aims to answer two main questions. The first is to determine the characteristics of the massive and dusty population of galaxies as a whole. The second is whether FIR-selected samples form a unique population of starbursting galaxies with distinct physical properties. We organize the paper as follows: In section 2 we present the data. Section 3 describes our approach for modeling galaxy SEDs. Section 4 presents our results. In section 5 we discuss our findings which we summarize in section 6. All magnitudes quoted are in the AB system. A Chabrier \citep{chabrier03} IMF is assumed throughout the paper. We assume a cosmology with $\Omega_\Lambda$ = 0.7, $\Omega_M$ = 0.3, and $H_0$ = 70 km s$^{-1}$Mpc$^{-1}$. 

\section{Observations}

\begin{figure*}[ht]
\includegraphics[width=\textwidth]{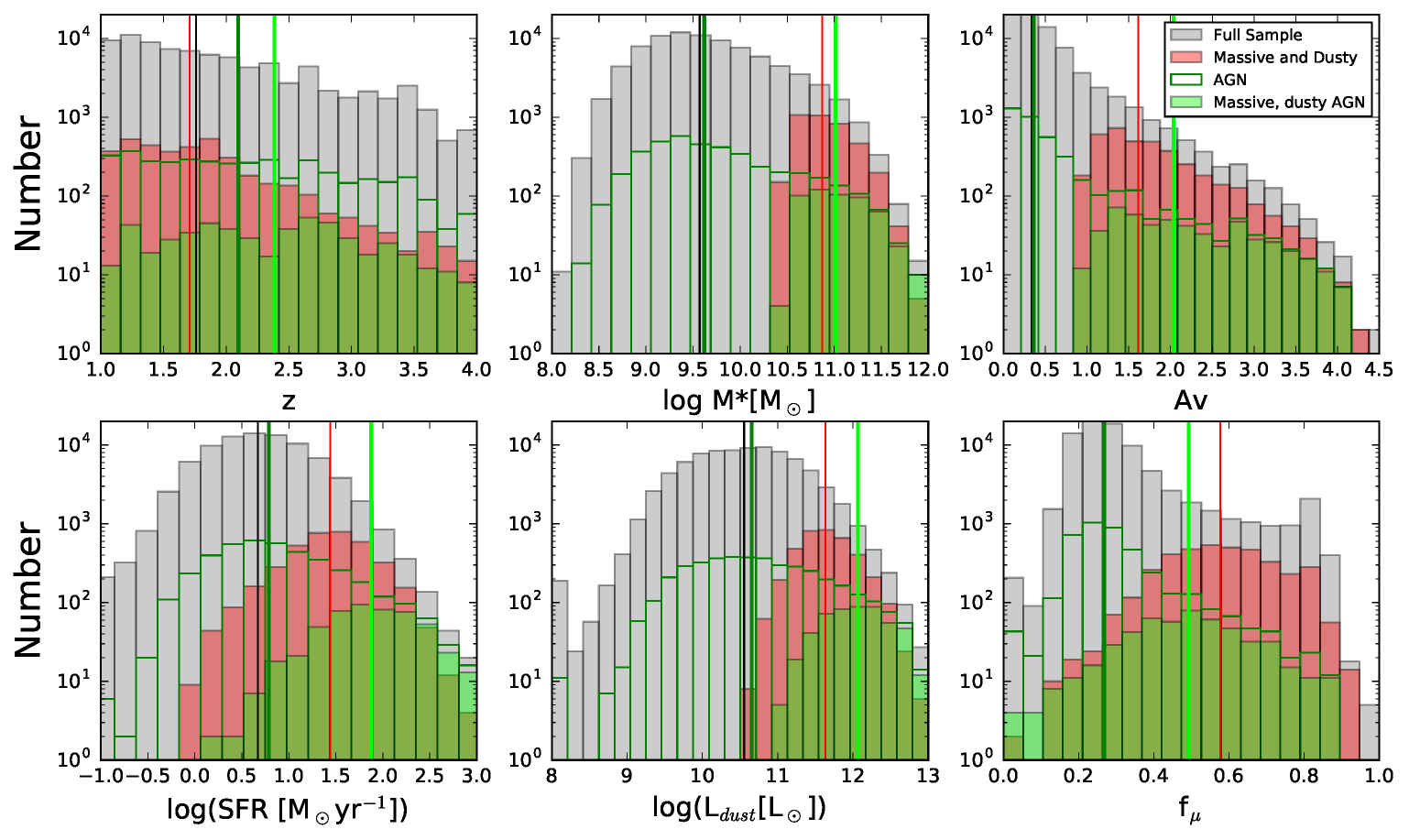}
\caption{Distributions of several key parameters for our sample. Gray and red bars correspond to the entire $1 \le z \le 4$ and the massive, dusty (see text) sample, respectively. The dark green histogram indicates AGN (see text) within the full sample, and the light green bars show AGN that have been removed from the massive and dusty sample. Medians for each of the distributions are indicated by vertical lines of corresponding color. \label{fig1}}
\end{figure*}

We select sources from the DR3 of the UltraVISTA photometric catalog. The DR3 catalog (Muzzin et al. in prep.) was constructed using the same procedure as in \citet{muzzin13a}. The relevant details for the current data release are as follows. The survey covers 0.7 deg$^2$ in the COSMOS field as deep stripes overlapping the DR1. The catalog includes photometry from the UV to \textit{Spitzer} 8$\micron$ with 49 bands. Sources are selected in the K$_S$-band, which has a point source 5$\sigma$ depth of 25.2 magnitudes. This depth allows us to be mass-complete at our adopted limit of log $(M_*/M_\odot) \ge 10.5$ for the final sample over our entire observed redshift range. In addition, \textit{Spitzer} MIPS 24$\micron$ observations \citep{lefloch09} are included as follows. Briefly, we assume that there are no color gradients in galaxies between the K$_S$-band and the IRAC and MIPS bands. The K$_S$-band is then used as a high-resolution template image to deblend the IRAC and MIPS photometry. Each source extracted from the K$_S$ image is convolved with a kernel derived from bright PSF stars in the K$_S$ and IRAC/MIPS images. The convolved galaxies are then fit as templates in the IRAC and MIPS bands with the total flux left as a free parameter. In this process, all objects in the image are fit simultaneously. Once the template fitting is converged, a "cleaned" image is produced for each object in the catalog by subtracting off all nearby sources \citep[for an example of this process see Figure 1 of][]{wuyts07}. Aperture photometry is then performed on the cleaned image of each source. For the IRAC (MIPS) channels the photometry is performed in a 3" (5") diameter aperture for each object. All UV-NIR fluxes are scaled to total using the ratio of total to aperture fluxes for the K$_S$-band and MIPS fluxes have been converted to total fluxes using an aperture correction of 3.7, as listed in the MIPS instrument handbook.

We supplement the UltraVISTA data with observations from the \textit{Herschel} PACS Evolutionary Probe  \citep[PEP;][]{lutz11} and \textit{Herschel} Multi-Tiered Extragalactic Survey \citep[HerMES;][]{oliver12, hurley16} surveys. The PEP survey covers most of the UltraVISTA footprint with the PACS 100$\micron$ and 160$\micron$ filters for which the 80\% completeness level is reached at 6.35 and 14.93 mJy (5$\sigma$ calculated from one sigma noise is 7.50 and 16.35 mJy), respectively. HerMES covers the area with the 250\micron, 350\micron, and 500$\micron$ filters at 5$\sigma$ depths of 15.9, 13.3, and 19.1 mJy, respectively. For both surveys we use the source catalogs extracted using MIPS 24$\micron$ priors which also use the \citet{lefloch09} data. For HerMES we use the most recent XID+ catalogs \citep{hurley16}. Sources with at least a 3$\sigma$ detection in any of the Herschel bands are matched to UltraVISTA sources within a matching radius of 1.5" provided the corresponding UltraVISTA source is detected in MIPS.

\section{SED Modeling}

Redshifts for the entire UltraVISTA catalog are determined using the photometric redshift code EAZY \citep{brammer08}. EAZY fits the spectral energy distribution (SED) by matching linear combinations of templates. The template set used for the DR3 is similar to that described in \citet{muzzin13a} for the DR1. Briefly, there is a set of templates derived from the PEGASE models \citep{fioc99}, a red template from the models of \citet{maraston05}, a post-starburst \citet{bc03} model, as well as a template to account for galaxies that are both old and dusty. For a detailed description of the fitting method, see \citet{brammer08}. We define the best redshift for each source as the spectroscopic redshift when available, or the peak of the redshift probability distribution from EAZY otherwise. We restrict our analysis to sources with $1 \le z \le 4$ and a $K_S$-band total magnitude less than 25. This leaves us with 85,286 sources. 

We use the high-z extension of MAGPHYS \citep{dacunha08, dacunha15} to model the UV-FIR photometry for all our sources.  We summarize the key components of the model here. Stellar emission is attenuated using the two component dust model of \citet{charlot00}. Young stars are attenuated by a dense birth cloud component which has a lifetime varying from 5 to 50 Myr as well as the ambient interstellar medium (ISM), whereas older stars only suffer attenuation by the ISM. The code generates a library of stochastic models of stellar populations at the redshift for each input source. These are combined with another random library of infrared spectra with the condition that any attenuated stellar emission must be accounted for by the matched infrared model. The code compares synthetic photometry of these models with the data to determine the best fit spectrum and provide the likelihood distributions of key physical parameters for the galaxy. We choose MAGPHYS due to this self-consistent way in which the entire SED is modeled such that energy absorbed by dust in the UV must be re-radiated in the infrared. This is important if we wish to consistently compare stellar population properties of a galaxy to its dust properties. For our estimates of galaxy physical properties we take the medians of the output probability distributions for each parameter. Corresponding errors are taken to be the maximum of the ($84^{th}-50^{th}$ percentile) and ($50^{th}-16^{th}$ percentile).

Since we wish to incorporate far-infrared (FIR) information for all of our sources consistently and many of our sources are not detected in any given FIR band, we adopt the following approach. We run MAGPHYS using directly observed fluxes, setting the \textit{Herschel} fluxes for any unmatched sources to non-detections for that band. We incorporate upper limits on the far-IR fluxes by using the depths of the PEP and HERMES surveys ($3\sigma$ upper limits of 4.5, 9.8 ,9.5, 8.1, 11.4 mJy for the 100, 160, 250, 350 and 500 $\micron$ bands, respectively). In Appendix A, we compare the MAGPHYS derived stellar and dust properties with and without the inclusion of the Herschel photometry for those sources detected in all five Herschel bands. Appendix A shows that quantities such as the stellar mass (M$_*$), the star-formation rate (SFR), dust extinction (A$_V$), infrared luminosity (L$_{dust}$), and the fraction of the L$_{dust}$ originating from the diffuse ISM rather than the birth cloud (f$_\mu$) are fairly robustly derived, whereas the dust temperature and the dust mass are only poorly constrained without the Hershel photometry. We do however stress that upper limit information is included in the modeling of all of our sources, and this contributes to better constrain the aforementioned quantities even when detections in the Hershel bands are not available.

When modeling broadband photometry, the assumptions of the underlying characteristics of the stellar populations can strongly influence the derived physical properties \citep[e.g.,][]{muzzin09, leja19a}. Specifically of concern to the present analysis are the potentially degenerate reddening due to higher metallicity and dust extinction as well as the effects of the assumed form of the star formation history. MAGPHYS allows the metallicity to vary from $0.2-2 Z_\odot$. \citet{muzzin09} show that the effect of different assumed metallicities on A$_V$ can be as large as 0.4 mag. Without spectroscopic information, it is not possible to break the degeneracy between metallicity and dust extinction. The star-formation history parametrization adopted by MAGPHYS is a continuous delayed exponential of the form: $\gamma^2 t e^{(-\gamma t)}$ where $\gamma = 1/\tau$ or the inverse of the star formation timescale. To account for stochasticity, star formation bursts of randomized magnitude and duration are superimposed on top of the model. This form, motivated by \citet{lee10} is meant to reflect the growing evidence that high-redshift galaxies should have rising star formation histories rather than the declining ones commonly assumed for low redshift galaxies \citep{Behroozi13, Pacifici13, Simha14}. The shape of the star formation history influences the derived stellar mass and SFR. Although the star-formation history is not constrained by modeling of the broadband photometry alone (e.g., Leja et al. 2018b), the flexible and comprehensive behavior of the SFH adopted in MAGPHYS helps reducing systematic effects introduced by too simplistic functional forms of the SFH (e.g., exponentially declining). Moreover, the inclusion of the Hershel photometry (either through detections or upper limits) ensures stronger constraints on the estimated SFRs compared to SFR estimated from the modeling of the UV-to-NIR photometry alone.

MAGPHYS outputs best fit attenuated and unattenuated SEDs for each source. The code calculates visual band extinction (A$_V$) by taking the difference in magnitudes obtained after integrating the SEDs over a rest-frame V-band filter. In order to construct our massive and dusty sample, we select sources that have redshifts $1 \le z \le 4$, M$_* \ge 10^{10.5}$M$_\odot$ and A$_V \ge 1$. This selects a total of 4,250 sources out of the 9,468 that meet only the mass and redshift cuts. A visual inspection of the best-fit SEDs resulted in the removal of an additional 274 sources. Table 1 shows the fraction of sources detected in MIPS and each of the \textit{Herschel} bands after AGN removal (see below). A total of 60\% of sources are detected at $3\sigma$ in MIPS, whereas $\sim 32\%$ sources are detected in both MIPS and at least one \textit{Herschel} band. This latter group of sources will be referred to our FIR-detected subsample throughout the paper.

\begin{table}[]
    \centering
    \begin{tabular}{c|c}
    Band     &  Detection Fraction (\%) \\
    \hline
    MIPS 24$\micron$     & 65 \\
    PACS 100$\micron$     & 4\\
    PACS 160$\micron$     & 4\\
    SPIRE 250$\micron$     & 35\\
    SPIRE  350$\micron$     & 22\\
    SPIRE 500$\micron$     & 9\\
    \end{tabular}
    \caption{Fraction of sources in our massive and dusty sample (see text) detected at 24$\micron$ and in each \textit{Herschel} band.}
    \label{Table 1}
\end{table}

\begin{figure*}[ht]
\includegraphics[width=\textwidth]{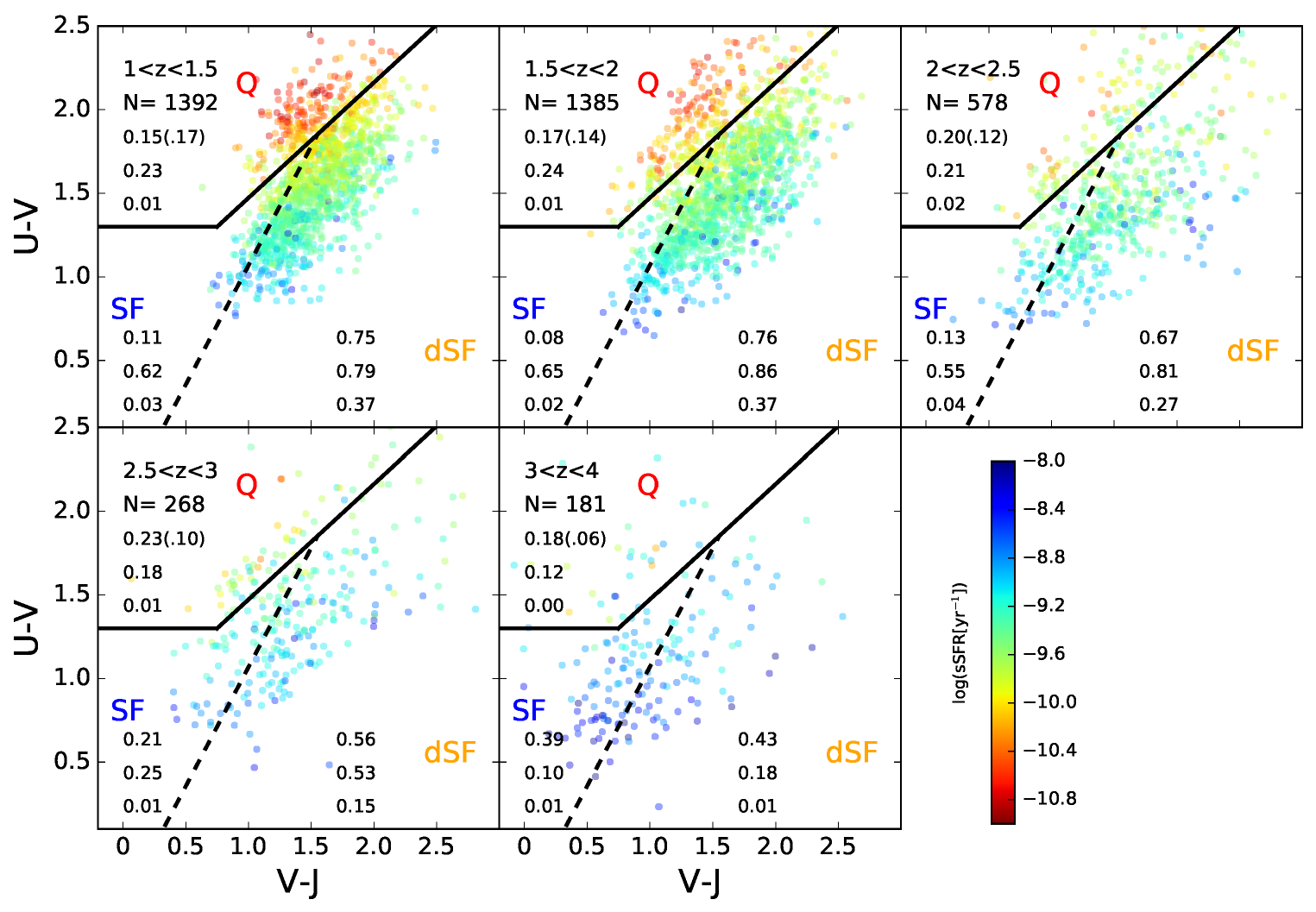}
\caption{UVJ diagram for the massive, dusty sources (see text) split into bins by redshift as indicated in each panel. The number of sources in each bin is also shown. Regions are labeled as quiescent (Q), non-dusty star-forming (SF), and dusty star-forming (dSF). For each region of the diagram we indicate the fraction of sources in that region, as well as the fractions of sources within that region which are detected by MIPS and \textit{Herschel} respectively. In the quiescent region, we also show in parenthesis the fraction of sources that would be identified as quiescent by a cut in sSFR at $10^{-10}$yr$^{-1}$. The dashed line corresponds to the criterion in \citet{martis16} to select "dusty" star-forming galaxies. Coloring of the points indicates the sSFR of each source determined by MAGPHYS. \label{fig2}}
\end{figure*}

Figure 1 shows the distributions of several parameters from the SED modeling for both the full $1 \le z \le 4$ sample (gray) and our massive, dusty selection (red) after AGN removal (see below). Other than the redshift, each of these parameters is the median value from the MAGPHYS SED fit. Medians of each of the distributions are indicated by vertical lines of corresponding color. The top three panels correspond to the parameters used to select our sample for analysis. Interestingly a massive and dusty selection does not appear to significantly alter the shape of the redshift distribution of our sources for $z \ge 1$. It is also worth noting that selecting massive galaxies already selects many of the most heavily obscured sources, as can be seen from the high degree of overlap of the red and gray histograms in the A$_V$ panel. The bottom three panels show the SFR, dust luminosity, and f$_\mu$ distributions. As mentioned above, MAGPHYS uses a two component dust model consisting of stellar birth clouds and diffuse ISM. The f$_\mu$ parameter indicates the fraction of the dust luminosity originating from the diffuse ISM. Selecting a massive subsample from a majority star-forming parent sample expectedly increases the median SFR in accordance with the observed correlation of stellar mass and SFR for star-forming galaxies, the so-called star-forming main sequence. The requirement of high extinction values unsurprisingly leads to a median dust luminosity over an order of magnitude higher than the general population. About 97\% of our massive and dusty selection can be classified as luminous infrared galaxies (LIRGs) with IR luminosities greater than $10^{11}$L$_\odot$, whereas about 17\% are ultra-luminous infrared galaxies (ULIRGS) defined by IR luminosities greater than $10^{12}$L$_\odot$. The massive and dusty selection thus makes up a large portion of the high-end tails of the SFR and dust luminosity distributions. It is interesting to note, however, that a number of highly star-forming sources do not meet our selection criteria. Finally, we observe much higher f$_\mu$ values for our massive and dusty sample than the parent sample, with median values of $\sim 0.6$ and $0.25$, respectively. This means that our sources have a higher fraction of their IR luminosity originating from the diffuse ISM. The implications of this finding will be discussed in detail below.

For completeness, in Appendix B, we show the distribution of mass-weighted stellar ages for the same samples mentioned above (see Fig. 17), as well as the comparison between the stellar age and sSFR, f$_\mu$, Z, and A$_V$ (see Fig. 18).

\subsection{AGN Removal}

MAGPHYS does not include models to account for the emission due to active galactic nuclei (AGN). A version that does include AGN is in preparation and has been applied to \citet{chang17}. The SFR values we observe for some sources may thus be due to contamination by the AGN. In order to remove any potential biases in our analysis, we identify potential AGN using several diagnostics. We test MIR colors using the criteria of \citet{donley12}, match X-ray sources from the \citet{marchesi16} catalog, and match radio sources from the \citet{smolcic17} catalog. For sources with a 3$\sigma$ detection in the SPIRE 250$\micron$-band, we also use the IR color-based templates from \citet{kirkpatrick15} to select sources for which the MIR emission is expected to be dominated by AGN. This corresponds to sources in region COLOR4 and above in a 250$\micron$-24$\micron$  and 8$\micron$-3.6$\micron$ color-color diagram (see their section 4.2). The properties of the sources meeting any of these criteria from our initial sample are shown as a dark green histogram in Figure 1. A total of 519 sources satisfy at least one of these criteria in addition to our massive and dusty selection and are removed from our sample. These are shown as a light green filled histogram in Figure 1. This leaves 3,457 sources for the final analysis. We briefly note that AGN candidates make up around 15\% of the massive and dusty selection before removal, suggesting a high incidence in obscured, high-redshift galaxies. Additionally, the majority are found via IR selections, supporting previous claims that optical/x-ray selections can miss heavily obscured AGN \citep{ballantyne06, tozzi06}.

\section{Results}
\subsection{Rest-Frame Colors}

\begin{figure*}[ht]
\includegraphics[width=\textwidth]{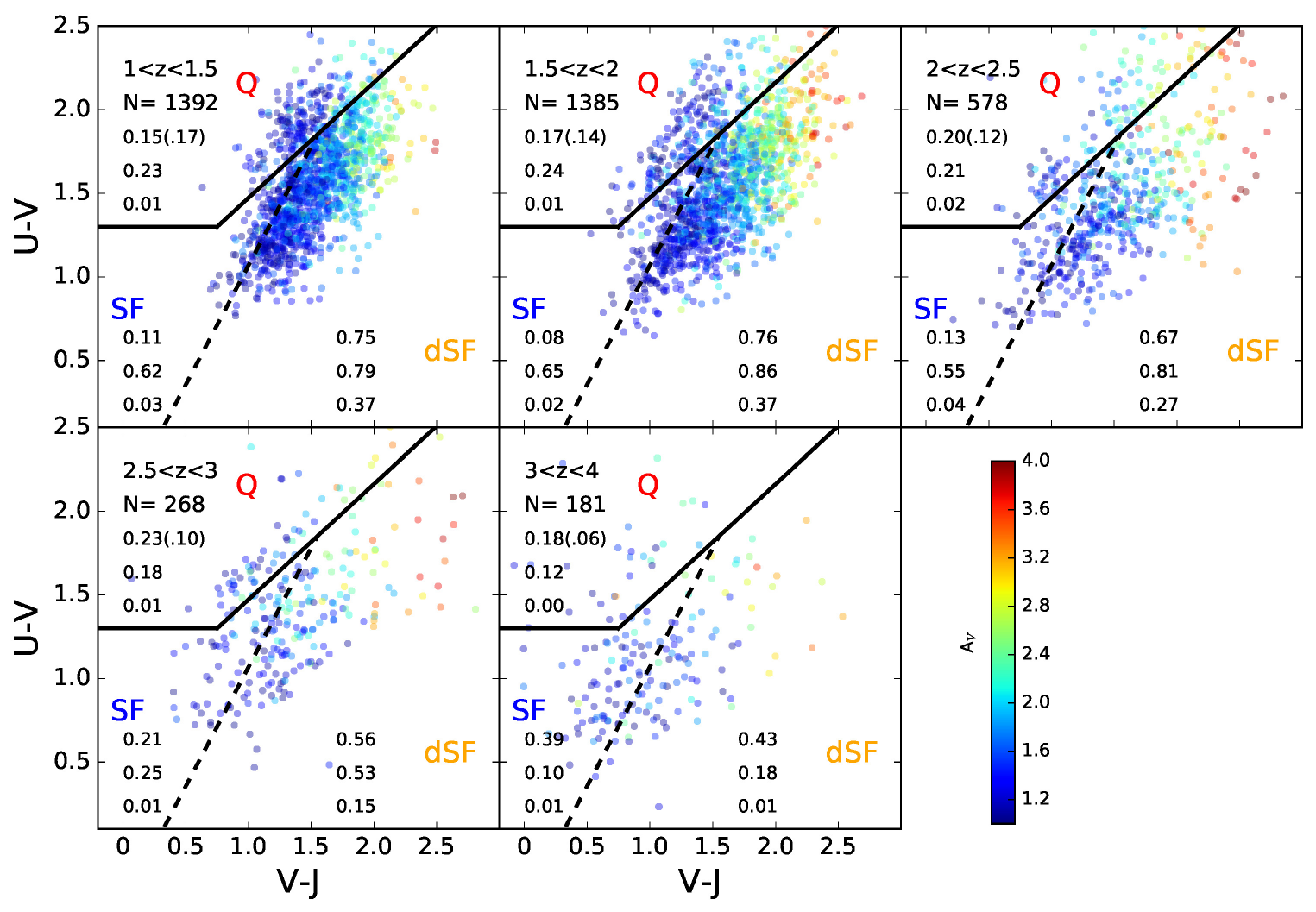}
\caption{UVJ diagram as in Figure 2.  Coloring now indicates A$_V$. \label{fig3}}
\end{figure*}

\begin{figure*}[ht]
\includegraphics[width=\textwidth]{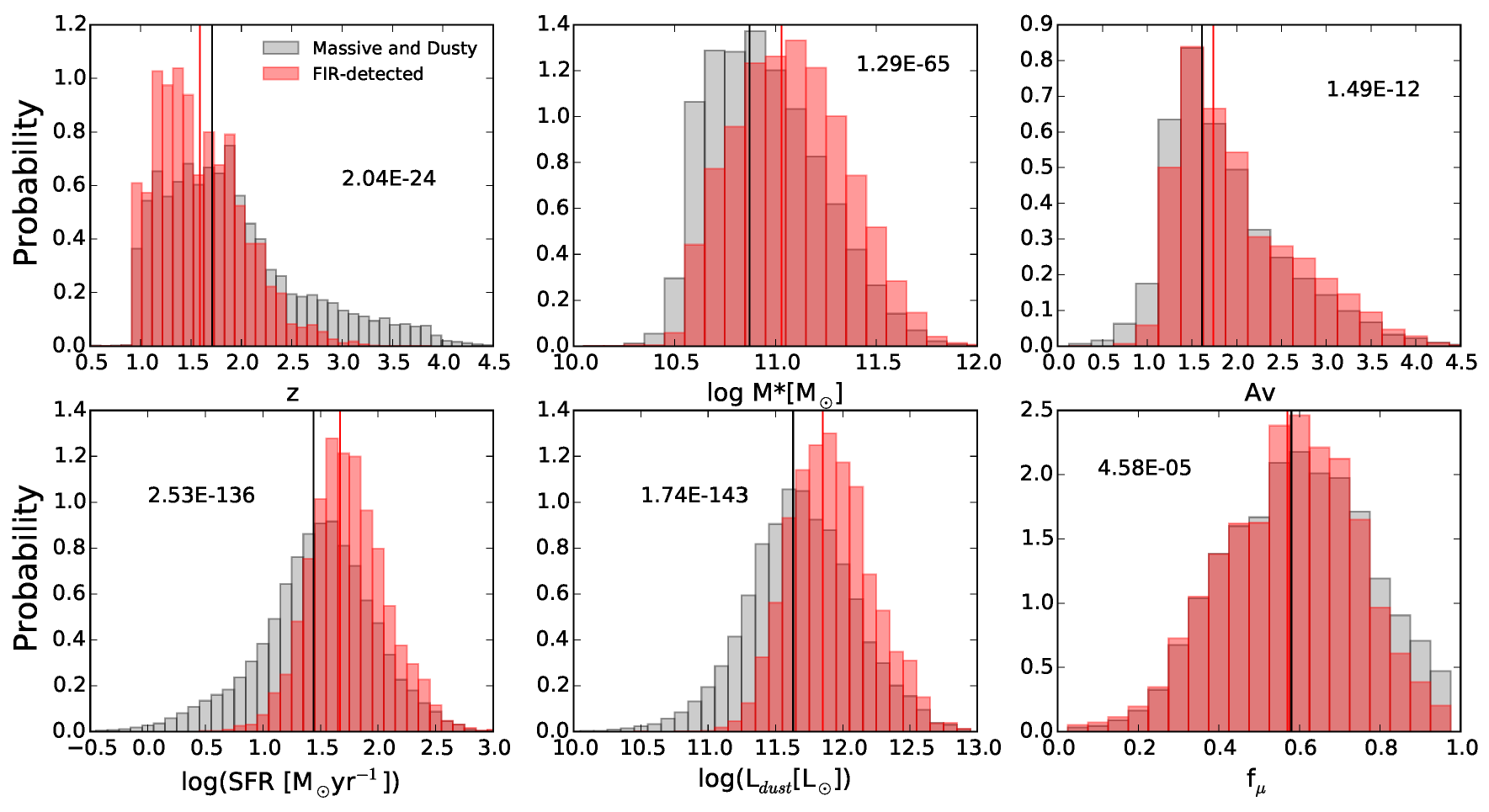}
\caption{Normalized stacked probability distributions of several key parameters for the FIR-detected subsample (red) compared to the full massive and dusty sample (gray). Medians of the distributions for the individual sources are indicated by vertical lines in corresponding colors. The p-value for a two sample KS test comparing the subsample distributions is also shown in each panel. \label{fig4}}
\end{figure*}

One of the primary properties we wish to investigate for our massive, dusty sample is the level of star-formation. Additionally, we wish to quantify the properties of star-forming and quiescent sources separately. The rest-frame $U-V$, $V-J$ color-color (UVJ) diagram has become a standard tool for accomplishing this selection due to its ability to distinguish reddening due to an aging stellar population from that caused by dust obscuration \citep{williams09, brammer09, whitaker12}. We calculate rest-frame UVJ magnitudes from the best-fit EAZY SED using the method of \citet{brammer11}. Figure 2 shows the UVJ diagram for our sample split into bins of redshift as indicated. Following \citet{martis16}, we define the quiescent population as satisfying 
\begin{equation}
(U-V) >  1.3 \textrm{ for }(V-J) < 0.75
\end{equation}
and
\begin{equation}
(U-V) >  0.69(V-J) + 0.7 \textrm{ for } (V-J) \ge 0.75
\end{equation}
Additionally, we show the division in the star-forming region of the diagram derived by \citet{martis16} to statistically correspond to A$_V \sim 1$. In that work galaxies satisfying 
\begin{equation}
(U-V) < 1.43(V-J) - 0.36.
\end{equation}
are designated as dusty star-forming galaxies. The division from \citet{martis16} is derived empirically by tracing the location of sources with A$_V \sim 1$ in the UVJ diagram similar to Figure 3 below. Due to differing star formation histories among galaxies, there is necessarily some cross-contamination between the non-dusty and dusty regions, but this division was shown to be statistically robust over a wide range in stellar mass up to $z \le 3$. 

In order to check the validity of the UVJ selection of star-forming and quiescent galaxies for our sample, we examine the distribution of SFRs determined by MAGPHYS in the UVJ diagram. The SFR given by MAGPHYS is averaged over a 10 Myr period. Figure 2 shows the derived specific star formation rate (sSFR) for each source via color coding. In each redshift bin we also indicate the fraction of sources in each of the three regions, the fraction of sources in that region with MIPS detections, and the fraction in that region with at least one \textit{Herschel} detection respectively. In the quiescent region, we also show in parenthesis the fraction of sources that would be identified as quiescent by a cut in sSFR at $10^{-10}$yr$^{-1}$. This value has been shown in previous work to agree well with UVJ color selections \citep[e.g.][]{wu18}. 

Up to $z \le 3$, the UVJ classification agrees relatively well with the specific sSFRs derived from MAGPHYS. Specifically, galaxies in the UVJ star-forming region typically have large values of the sSFR, while galaxies in the UVJ quiescent region are mostly characterized by small values of sSFR. We however note that there are some disagreeing sources in the quiescent UVJ region with MAGPHYS sSFR $>10^{-10}$ yr$^{-1}$. We also notice that the quiescent region of the UVJ diagram becomes increasingly populated with objects of higher sSFR with increasing redshift. We note that many of these higher sSFR objects in the quiescent region lie near the boundary and up to $z \sim 2$ exhibit sSFRs intermediate between the truly quiescent sources and sources with similar colors in the star-forming region, so that these could likely be galaxies in the process of quenching or transitioning. For our highest redshift bin, MAGPHYS assigns almost uniformly high SFRs and sources are concentrated in the non-dusty star-forming region, so in this regime the UVJ classification may be less robust. Alternatively, the apparent disagreement may be evidence for spatial separation of the regions generating the rest-frame UV-NIR emission, which determines UVJ colors, and the IR emission, which constrains the SFR, for these sources. Examples of such cases are presented in \citet{elbaz18, schreiber18}. We find the fraction of quiescent sources selected by sSFR to be qualitatively similar to those derived from UVJ colors. For all but the $1<z<1.5$ bin, fewer sources are identified as quiescent by the sSFR selection. The largest discrepancy occurs in our highest redshift bin where nearly all sources have sSFR $>10^{-10}$ yr$^{-1}$.

Previous work has shown systematic trends in A$_V$ across the UVJ diagram \citep{martis16, fang17}. Figure 3 shows the distribution of A$_V$ for each of our sources across the UVJ diagram. We find similar trends here, with extinction values increasing toward the upper right of the diagram. Up to $z \sim 3$ we find that most of our star-forming sources satisfy the dusty star-forming selection of \citet{martis16} despite differences in modelling and the assumption of a different dust law. This supports the robustness of the \citet{martis16} color cut to select heavily-obscured sources. Specifically, UVJ-defined dusty star-forming galaxies represent $\sim 70\%-80\%$ of the sample of massive and dusty galaxies out to $z=2.5$, with small contribution ($\sim 5\%-10\%$) from relatively unobscured star-forming galaxies. At $2.5<z<3$, we see evidence for a decline in the fraction of dusty star-forming galaxies ($\sim 60\%$) with an increasing role of the relatively unobscured star-forming galaxies ($\sim 20\%$). In our highest redshift bin, the largest fraction of our sources lie in the non-dusty star-forming region. Due to the dependence of extinction on the dust geometry, the increasing fraction of sources with high extinction in the non-dusty star-forming region of the UVJ  at $z \gtrsim 3$ may reflect changing geometry of star-forming regions or different dust properties at these redshifts. As further support for our UVJ classification we find that our FIR-detected sources lie almost exclusively in the dusty star-forming region, where they make up about a third of the sources in that region up to $z \sim 2.5$ (Figures 2,3).

In the local universe massive, quiescent galaxies contain very little gas and dust \citep[e.g.][]{smith12}, so the number of UVJ quiescent sources that satisfy our selection may be surprising. Recent work \citep{gobat18} has shown that this may not be the case for their quenched progenitors at higher redshift. UVJ quiescent sources consistently account for $\sim 15-20\%$ of the selected sample across the observed range in redshift. Additionally, previous work has shown that a substantial fraction of quiescent galaxies at these redshifts are detected with MIPS at 24$\micron$. \citet{fumagalli14} observe a $\sim 25\%$ detection rate for quiescent galaxies in their rest-frame optical-selected sample. This fraction is about the same as the fraction of quiescent galaxies from our full sample which meet the $A_V \ge 1$ criterion.

\subsection{Comparison to FIR-Selected Populations}

Here, we first compare the properties of sources which are detected in at least one \textit{Herschel} band at 3$\sigma$, which we remind readers also requires a MIPS detection, to the properties of the full sample. Figure 4 shows the normalized stacked probability distributions of the same parameters as in Figure 1 for these two populations. Distributions for the FIR-detected sources are shown in red, whereas the full sample of massive and dusty galaxies is shown in gray. Vertical lines of corresponding colors indicate the medians of the distributions of the individual values adopted for each source. We also show the p-value obtained when performing a two sample Kolmogorov–Smirnov (KS) test on the detected and undetected subsamples. This determines if the two subsamples are consistent with being drawn from the same parent distribution. The top row shows the distributions for parameters that were used to select the massive, dusty sample. The FIR-detected subsample tends toward slightly higher stellar masses and A$_V$, but lower redshifts. The FIR-detected subsample distributions also lie at higher SFRs and dust luminosities. This is to be expected since a selection in FIR flux is essentially a selection in obscured star formation at a given redshift and the FIR flux directly constrains the dust luminosity. Interestingly, the distributions of A$_V$ for the two subsamples are comparatively similar according to the KS test. The f$_\mu$ parameter varies the least of those investigated between the subsamples, with FIR-detected sources marginally tending toward lower f$_\mu$. The KS tests shows that the probability that the distributions of the sources with at least one Herschel detection are drawn from the same parent population as the whole population of dusty and massive galaxies is negligible.

 \begin{figure*}[ht]
\includegraphics[width=\textwidth]{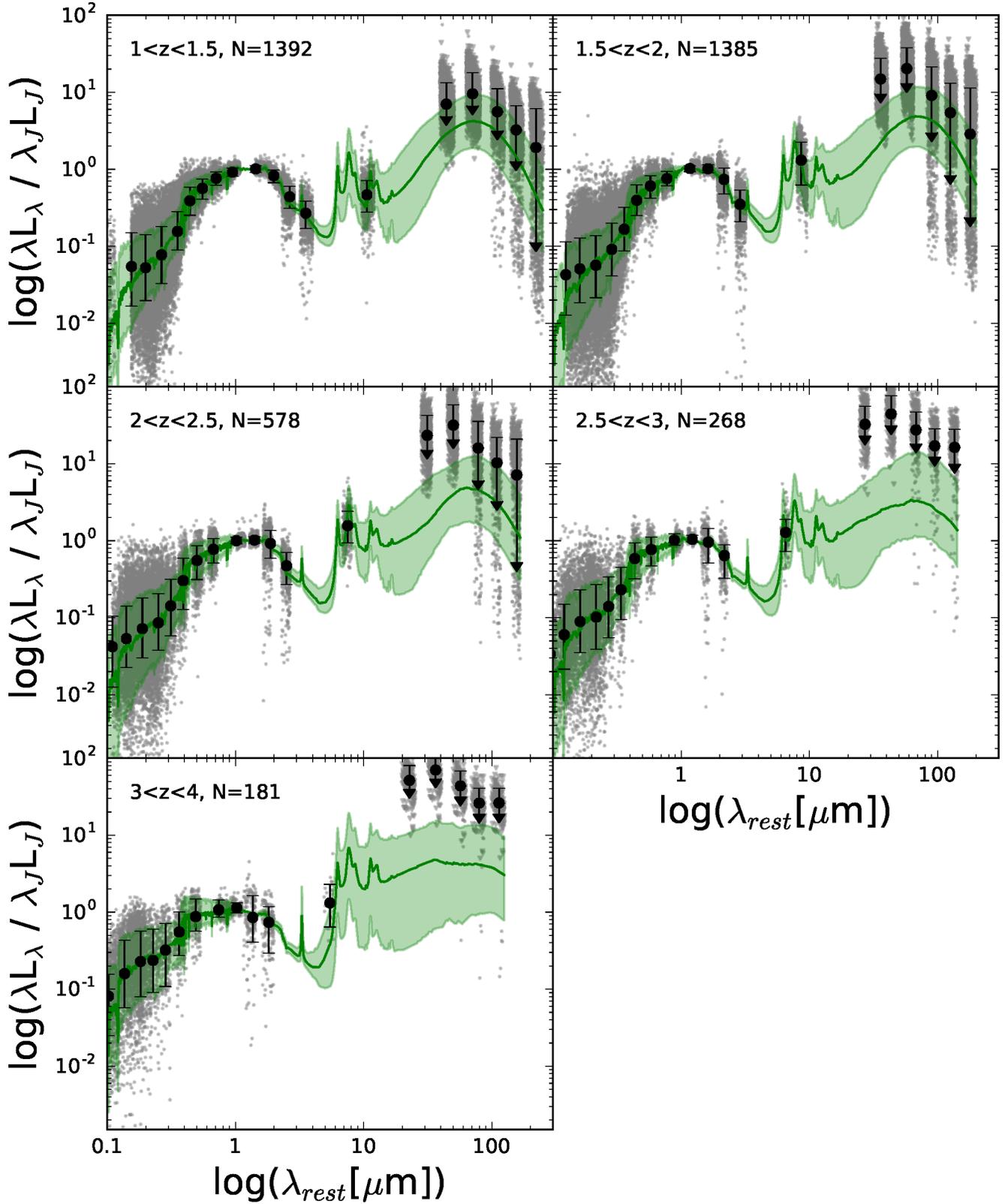}
\caption[width=\textwidth]{Median rest-frame SEDs for the entire sample with each panel corresponding to a different bin in redshift as labeled. Small gray points indicate individual observations. For the \textit{Herschel} bands, downward triangles indicate $3\sigma$ limits for undetected sources. Large black points with error bars show the median stacked observations and $15^{th}$ to $85^{th}$ percentiles of the distributions, including the $3\sigma$ limits. Green lines indicate the median stacked MAGPHYS models for the corresponding sources while shaded regions show the $15^{th}$ and $85^{th}$ percentiles of the model distributions. Model and observed fluxes are normalized to the rest-frame J-band of the model. \label{fig1}}
\end{figure*}
 
\subsection{Median SEDs of Massive, Dusty Galaxies}

\begin{figure*}[ht]
\includegraphics[width=\textwidth]{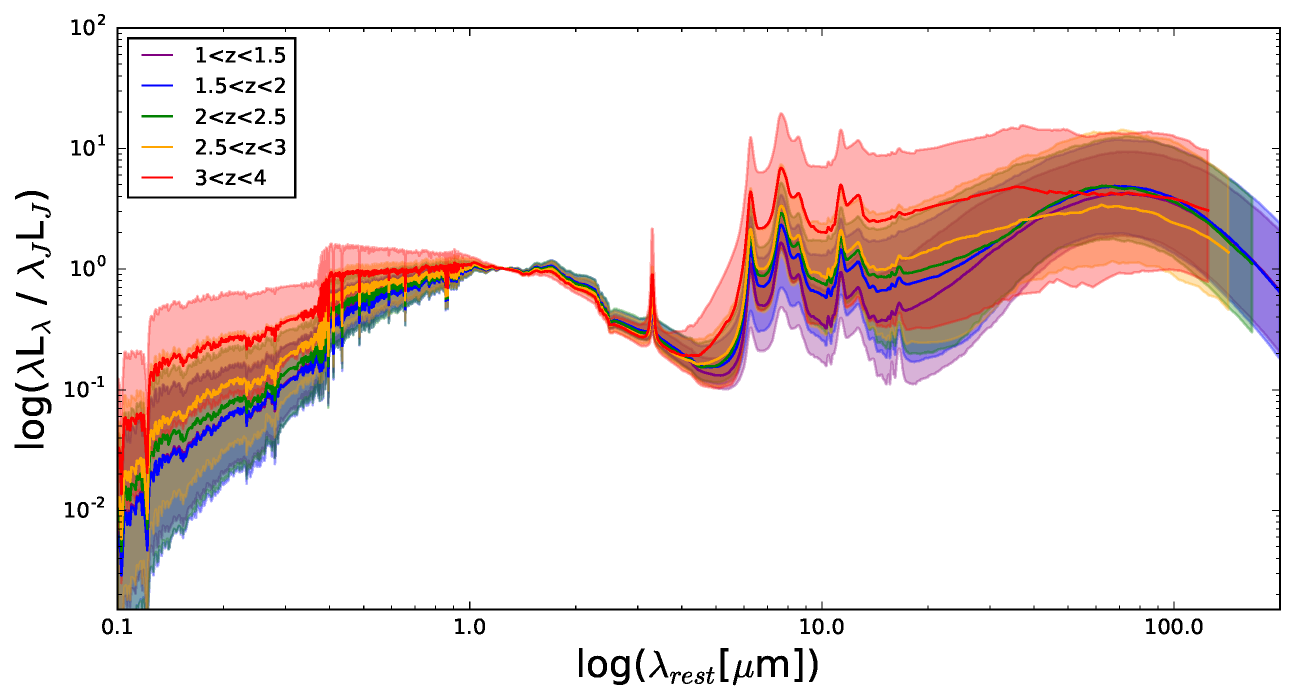}
\caption[width=\textwidth]{Median model SEDs for the entire sample split into bins of redshift as indicated by the coloring. Shaded regions show the $15^{th}$ and $85^{th}$ percentiles of the distributions. Fluxes are normalized to the rest-frame $J$-band.  \label{fig1}}
\end{figure*}

\begin{figure*}[h!]
\includegraphics[width=\textwidth]{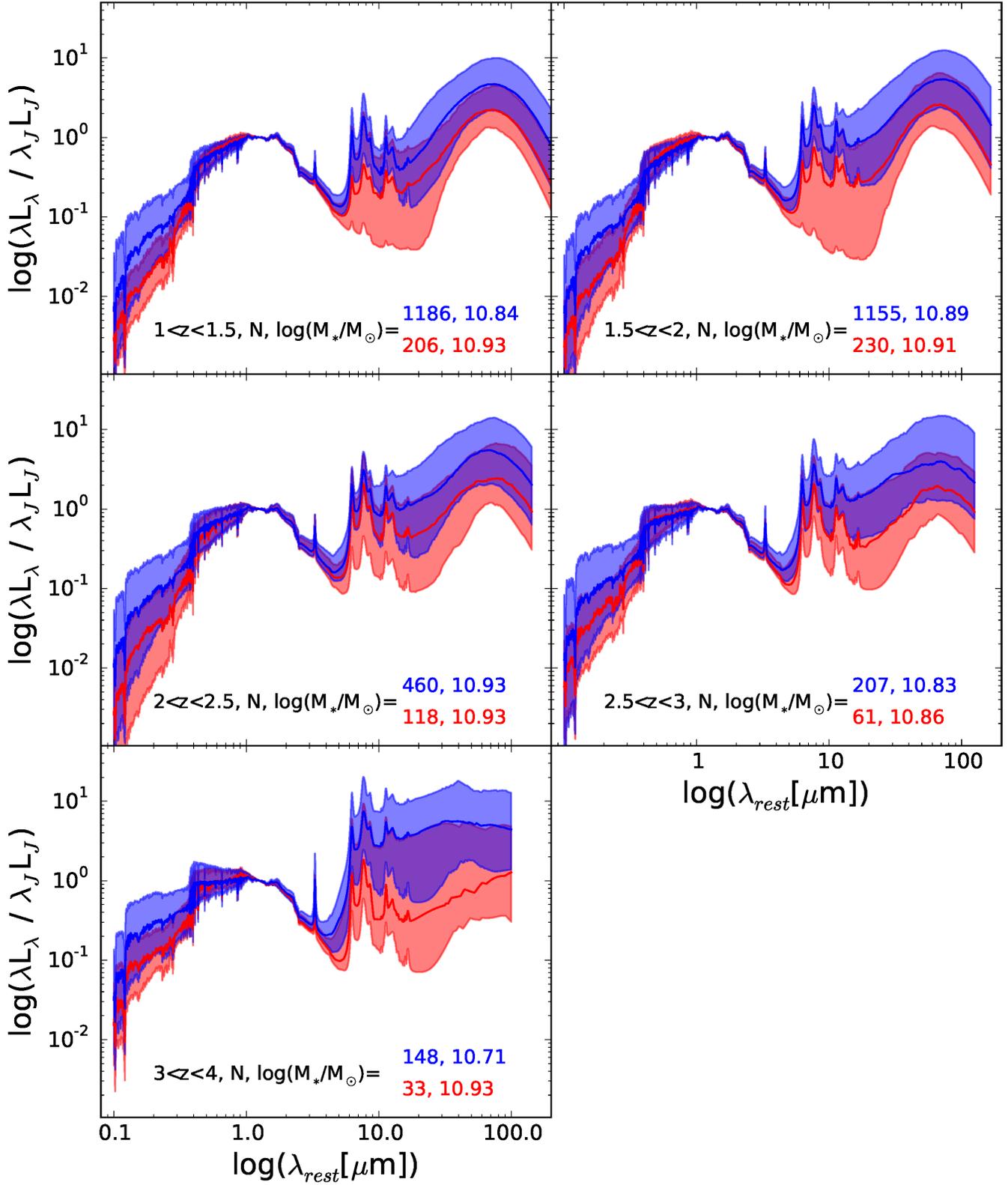}
\caption[width=\textwidth]{Median model SEDs for the entire sample with each panel corresponding to a different bin in redshift as labeled. Star-forming galaxies are shown in blue and quiescent galaxies in red, with the corresponding number and median stellar mass in each bin indicated. Shaded regions show the $15^{th}$ and $85^{th}$ percentiles of the distributions. Fluxes are normalized to the rest-frame J-band. \label{fig1}}
\end{figure*}

In order to better understand the evolution of this class of galaxies, we examine the median rest-frame SEDs from our sample in bins of redshift. The large black points in Figure 5 show the median stacked observations, individually shown with small gray points, for each of the redshift bins as labeled. Error bars indicate the $15^{th}$ and $85^{th}$ percentiles of the distributions. For the Herschel bands, we remind the reader that many of our sources are undetected in a given band. We have indicated this by plotting the $3\sigma$ upper limits as downward triangles for individual sources and by using downward arrows to show the locations of the $15^{th}$ percentiles of these distributions. Additionally, for sources that are only detected in one or two of the three SPIRE bands, the xid+ source extraction method provides estimation of fluxes for the undetected bands. These measurements represent the low fluxes for the SPIRE bands. The green curve shows the stacked MAGPHYS models for the corresponding sources, with the green shaded area representing the $15^{th}$ to $85^{th}$ percentiles of the model distributions. Fluxes are normalized to the rest-frame $J$-band of the model. From the UV-NIR the stacked models closely follow the shape and spread of the stacked observations. Figure 5 shows that the $3\sigma$ upper limits for many sources in the \textit{Herschel} bands lie well above the median values of the models, particularly for the PACS bands and at the highest redshifts. If we were to fit the FIR region of the SED independently, these high upper limits could pose a problem for the derivation of physical properties for our sample. Instead, the energy balance required by the MAGPHYS SED modeling constrains the FIR SED with both the observed \textit{Herschel} photometry and the inferred obscuration properties from the UV-optical region of the SED. 
%It appears as though the models underestimate the observations in the FIR, particularly for the PACS bands, but it should be recalled that a relatively small fraction of our sources are detected in PACS. It is expected then that sources with fluxes high enough to be detected would lie above the median fluxes of the rest of the sample. 

\begin{figure*}[ht]
\includegraphics[width=\textwidth]{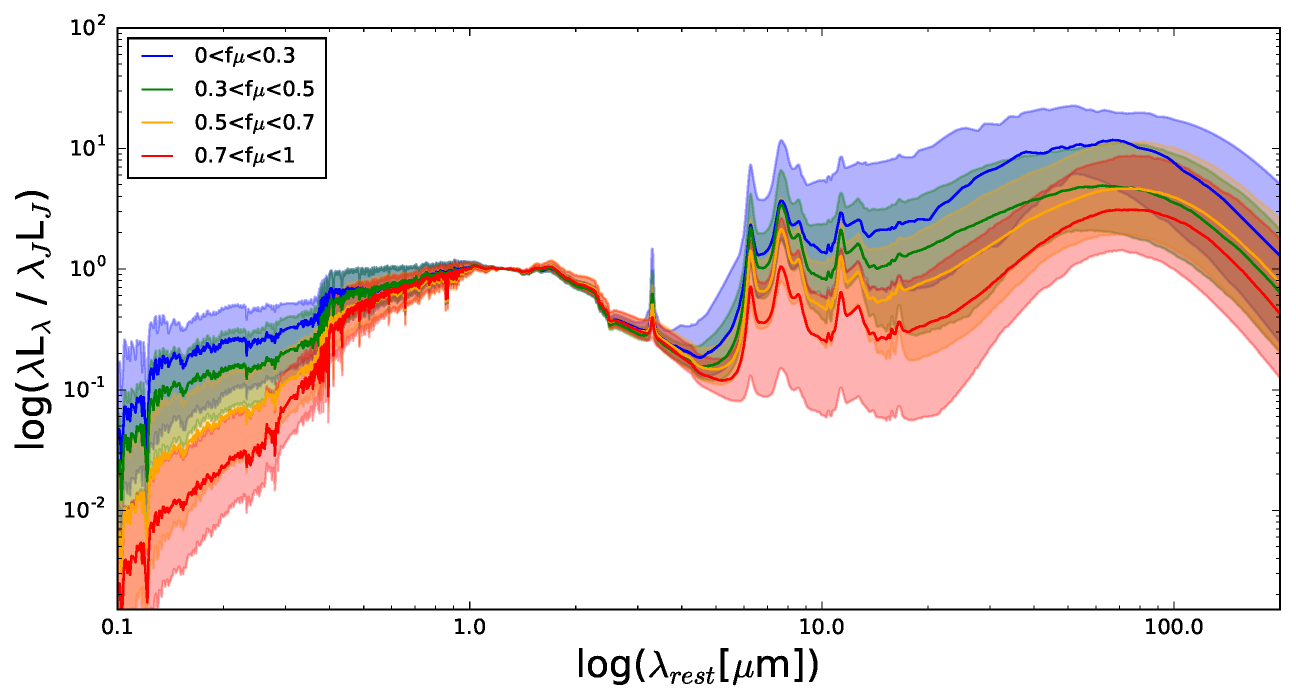}
\caption[width=\textwidth]{Median model SEDs for the entire sample split into bins of $f_\mu$ as indicated by the coloring. Shaded regions show the $15^{th}$ and $85^{th}$ percentiles of the distributions. Fluxes are normalized to the rest-frame J-band. \label{fig1}}
\end{figure*}

To make comparison of the median SEDs at different redshifts easier, Figure 6 shows the model SEDs of all galaxies from Figure 5 together, with redshift indicated by color. Shaded regions indicate the $15^{th}$ and $85^{th}$ percentiles of the distributions. Fluxes are normalized to the rest-frame J-band. The first thing to notice is the presence of a clear, mostly monotonic evolution of the median SEDs across the entire observed redshift range. In the UV-optical wavelengths the strength of the continuum relative to the NIR increases with redshift. This coincides with increasingly blue UV slopes with increasing redshift. MIR emission features near 10 $\micron$ likely to be dominated by polycyclic aromatic hydrocarbons (PAHs) are strongest in the highest redshift bin. Dust emission in the FIR relative to the NIR is also strongest in the highest redshift bin ($3<z<4$), and weakest in the smallest targeted redshift bin ($1.0<z<1.5$), although there does not appear to be much evolution in the contribution of this part of the spectrum at intermediate redshifts (i.e., $1.5<z<3$). The peak of the FIR emission occurs at the longest wavelength at $1.0<z<1.5$, and the shortest wavelength in the $3<z<4$ bin. Again, evolution within the intermediate redshift range appears to be minor. Thus there appears to be tentative evidence for a shift to increasing dust temperature with redshift for this population, although the large scatter in the SED shapes make this an uncertain result. We also see a trend toward a flattening of the FIR peak with increasing redshift. The depth of our $K_S$-band observations used for source extraction allows us to be mass-complete over the entire redshift range, so the evolution we see here should not be affected by selection bias. 

We also examine the rest-frame SEDs of star-forming and quiescent galaxies, as determined by UVJ colors, separately. We note that the designation of star-forming here and throughout the rest of the paper includes both the dusty and non-dusty star-forming regions of the UVJ diagram. Figure 7 compares the median star-forming, shown in blue, and quiescent, shown in red, galaxy SED for each of the redshift bins indicated. The number of star-forming and quiescent galaxies in each bin as well as the median log stellar mass for each selection are shown in blue and red, respectively. Again, shaded regions represent the $15^{th}$ and $85^{th}$ percentiles and fluxes are normalized to the rest-frame J-band. The median stellar mass for star-forming and quiescent sources in each bin is within 0.1 dex for all but the highest redshift bin, so differences in stellar mass should not complicate the comparison here. Unsurprisingly, even obscured star-forming galaxies emit more strongly in the UV at all redshifts, although the scatter here shows considerable overlap between the two populations in this regime. See the discussion section for possible interpretations. In most redshift bins, star-forming galaxies emit a stronger dust continuum that is also shifted to shorter wavelengths, suggesting higher dust temperatures. However, the scatter shows that these two populations again overlap to a large degree in this regime. The difference in cold dust emission between star-forming and quiescent galaxies appears to decrease with cosmic time, such that by our lowest redshift bin the median SEDs near the IR peak nearly overlap. We also see that the flattening of the FIR peak observed with increasing redshift in Figure 6 is due only to the star-forming segment of the sample. At all redshifts, star-forming galaxies also exhibit stronger MIR emission features. Interestingly, this distinction is weaker from $2.5 < z < 3$. Additionally, as can be seen in Figure 2, the sources classified as UVJ quiescent still have non-negligible SFRs from MAPGHYS, potentially indicating a shortcoming of the UVJ diagram for this redshift regime, at least when considering highly obscured sources.  A robust resolution of this issue would require a spectroscopic confirmation of the stellar and dust properties with the Atacama Large Millimeter Array (\textit{ALMA}) and the James Webb Space Telescope (\textit{JWST}). In the Appendix we show results corresponding to Figure 7 for a sSFR definition of quiescence.

\begin{figure}[h]
\includegraphics[width=\columnwidth]{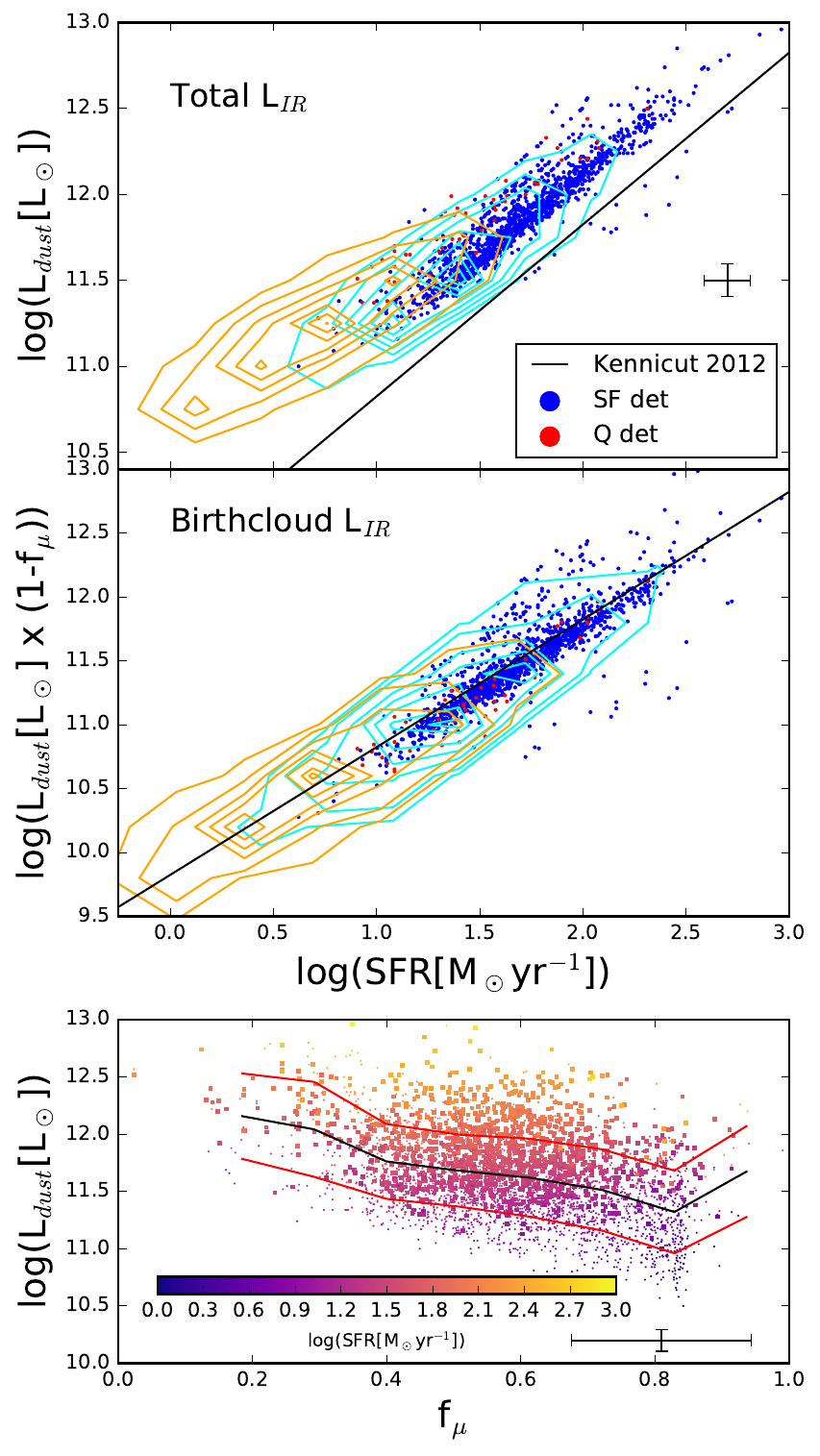}
\caption{Top: Total infrared luminosity as a function of SFR. UVJ star-forming sources are shown with cyan contours and UVJ quiescent with orange. Contours are placed at the 5$^{th}$, 15$^{th}$, 35$^{th}$, 50$^{th}$, 65$^{th}$, 85$^{th}$, and 95$^{th}$ percentile levels. Colored points indicate \textit{Herschel} detections for each type. The black line indicates the \citet{kennicutt98} relation scaled to a Chabrier IMF. Middle: As above, now showing infrared luminosity only due to emission by stellar birth clouds (see text). Bottom: Total infrared emission as a function of the fraction of dust emission arising from diffuse ISM, with color indicating SFR. Large points indicate \textit{Herschel} detections. The running median and one sigma scatter are shown by black and red curves, respectively. The crosses represent the median uncertainty in their respective panels. \label{fig1}}
\end{figure}

Lastly, we compare median SEDs in bins of the $f_\mu$ output from MAGPHYS in Figure 8. This quantity denotes the fraction of infrared luminosity contributed the diffuse ISM component of the dust. We remind the reader that MAGPHYS utilizes \citet{charlot00} dust model, which assumes two dust components, one for stellar birth clouds and one for the diffuse ISM. Birth clouds are heated by newly formed stars, so their emission traces SFR. ISM dust is heated by both young and evolved stars, and so more closely traces the stellar mass. A value of $f_\mu$=0 would thus correspond to all of a galaxy's dust emission arising from birth clouds, while $f_\mu$=1 would correspond to all of the dust emission originating from the ISM. As will be seen in Figures 9 and 11 below, $f_\mu$ anti-correlates with the sSFR in the galaxy in addition to weakly correlating with L$_{\rm IR}$. We observe a clear dependence on $f_\mu$ for the SED shape. In the UV, increasing values of $f_\mu$ correspond to a decreasing contribution of the UV to the overall luminosity as well as redder UV slopes. In the MIR, increasing $f_\mu$ corresponds to weaker emission features. The FIR dust continuum decreases with increasing $f_\mu$ in addition to shifting to longer wavelengths, suggesting lower dust temperatures. These trends are all in line with what one would expect given an anti-correlation of $f_\mu$ and SFR.

\subsection{Trends with Star Formation Rate}

\begin{figure}[h]
\includegraphics[width=\columnwidth]{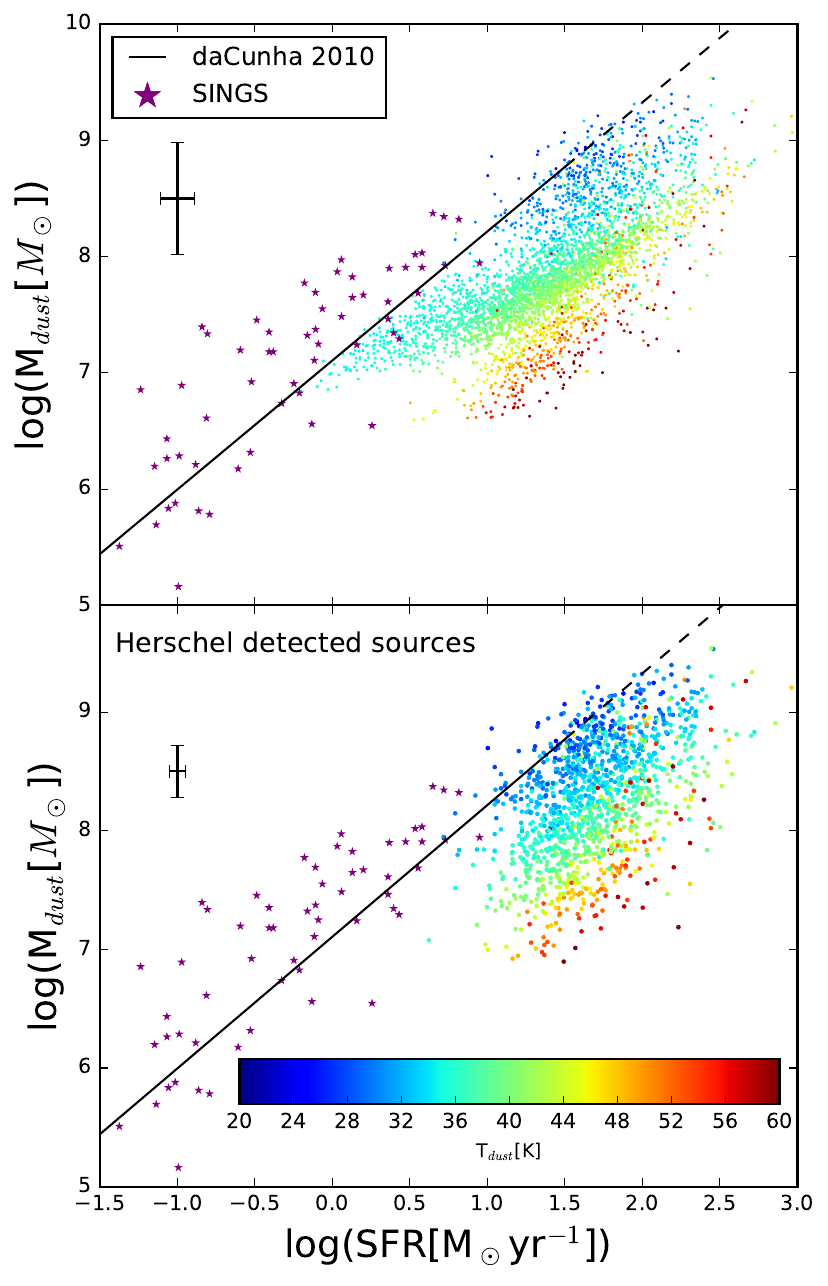}
\caption{Top: Dust mass versus SFR with dust temperature indicated by coloring. The black line shows the relation derived by \citet{dacunha10a}, with the dashed region showing the extrapolation to cover the same parameter space as our data. We also show the measurements from the local SINGS sample \citep{kennicutt03} for comparison. Bottom: as above, but only showing sources detected by \textit{Herschel}. The typical uncertainty in SFR and M$_{dust}$ is plotted in both panels. \label{fig1}}
\end{figure}

\begin{figure}[h]
\includegraphics[width=\columnwidth]{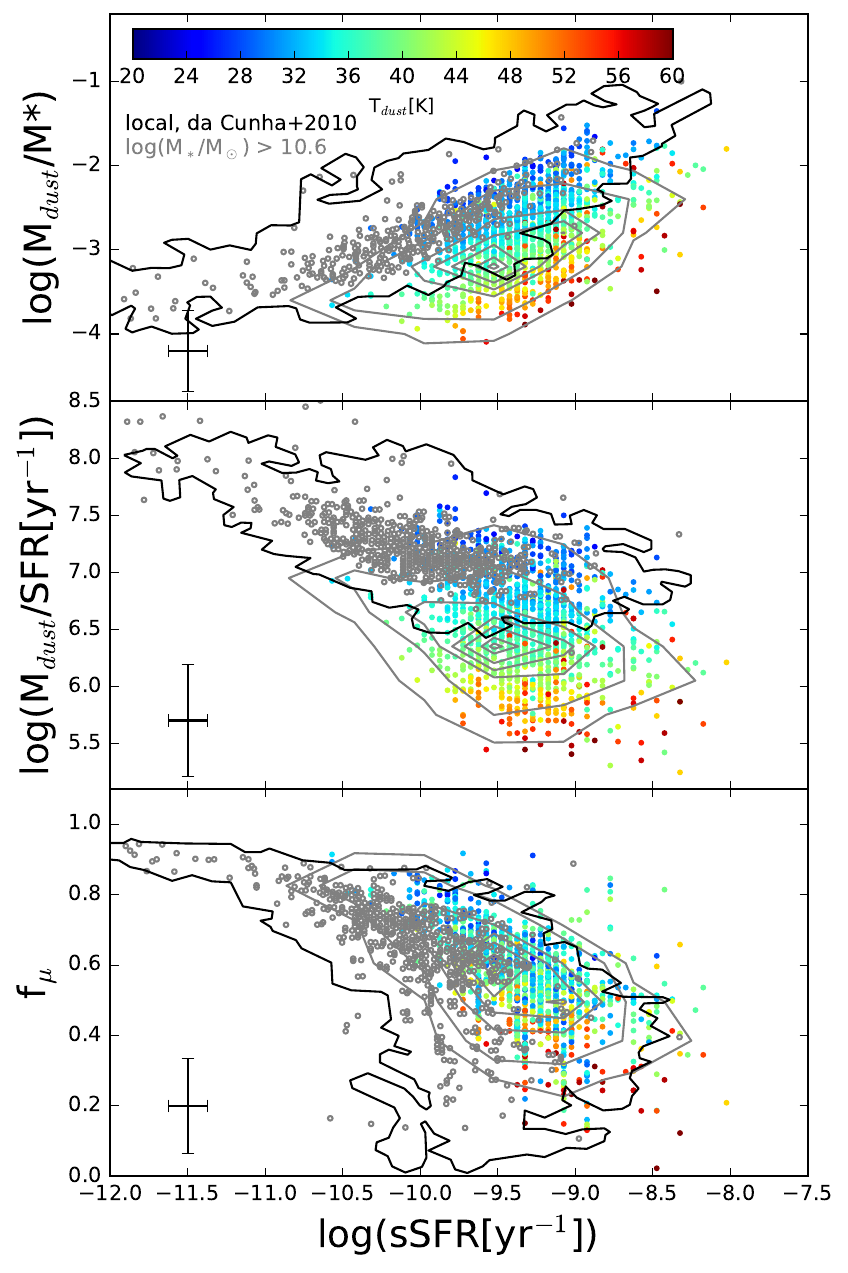}
\caption[width=\columnwidth]{Several ISM diagnostics as a function of sSFR. Our full sample is shown with gray contours while colored points indicate \textit{Herschel} detections. Contours are placed at the 5$^{th}$, 15$^{th}$, 35$^{th}$, 50$^{th}$, 65$^{th}$, 85$^{th}$, and 95$^{th}$ percentile levels. Coloring indicates dust temperature in all panels. Top: Dust to stellar mass ratio. Middle: Dust mass divided by SFR. Bottom: f$_\mu$, fraction of infrared luminosity contributed by dust in the diffuse ISM. The distribution of the full sample from \citet{dacunha10a} is shown with the black contour, whereas their subsample of massive (log(M$_*$/M$_\odot$)$>10.6$) galaxies is shown with gray circles. \label{fig10}}
\end{figure}

We have seen that a selection of massive and dusty galaxies at high redshift includes objects with a range of SED shapes and star formation activity. Additionally, we have seen that requiring a FIR detection leads to a bias in stellar mass, SFR, and dust extinction. Here, we more closely examine the physical properties of our sample. 

The total infrared luminosity (L$_{\rm IR}$) is known to be a good tracer of the SFR, with the canonical relation given by \citet{kennicutt98} \citep[updated][]{kennicutt12}. This relation is calibrated on optically-thick starbursts and so assumes all of the emission related to star formation is reradiated in the infrared. In the top panel of Figure 9 we show L$_{\rm IR}$ as a function of SFR, both of which are calculated from the  MAGPHYS SED fitting. We show the distributions of all quiescent and star-forming galaxies with orange and cyan contours respectively. Individual points indicate detections with \textit{Herschel}. The black line shows the Kennicutt relation scaled to a Chabrier IMF. We see that our data generally follow the slope of the relation, with sources at large star-formation rates showing a relatively small scatter, while the scatter increases significantly with decreasing SFRs. However, for most of the sources there appears to be an offset to larger IR luminosities with respect to the Kennicutt relation. In fact, only a small number of sources actually reside on the Kennicutt relation. Both the increasing scatter at lower star-formation rates and the overall offset are evidence for the presence of an important source of dust heating and IR emission other than star formation dominating in this regime. Indeed, if we look at star-forming and quiescent galaxies separately, we see this to be the case. The star-forming galaxies form a much tighter relation and lie closer to the Kennicutt relation. In contrast, the quiescent galaxies form a cloud at lower SFRs that is offset well above the relation. This is broadly consistent with the results of \citet{dacunha15} who find a similar offset for galaxies with older ages in a sample of SMGs observed with \textit{ALMA}. 

Our modeling allows us to more closely investigate this relation by considering the sources of the IR emission. The middle panel of Figure 9 shows L$_{\rm IR}$ scaled by  $1 - f_\mu$ versus SFR. This scaling means that we are only taking into account IR emission which according to the MAGPHYS fit is due to stellar birth clouds. We see that the UVJ star-forming sources now follow the Kennicutt line. Additionally, the majority of UVJ quiescent sources now continue this same relation, indicating that warm dust emission from birth clouds can trace even residual levels of star formation. 

Lastly, in order to better show the dependence of L$_{\rm IR}$ on both f$_\mu$ and SFR, we show in the bottom panel of Figure 9 L$_{\rm IR}$ as a function of f$_\mu$, that is the fraction of L$_{\rm IR}$ arising from diffuse ISM. Coloring indicates SFR while large points again indicate \textit{Herschel} detections. Black and red curves indicate the running median and one sigma scatter, respectively. We find that in addition to the expected relation between SFR and L$_{\rm IR}$, the total L$_{\rm IR}$ also correlates with the fraction of dust emission generated in stellar birth clouds, that is decreasing $f_\mu$. In our sample of massive, dusty galaxies, the median L$_{\rm IR}$ increases from log($L_{dust}/L_\odot) \sim 11.4$  when dominated by ISM emission to log($L_{dust}/L_\odot) \sim 12.4$ when dominated by birth cloud emission.

\begin{figure}[h]
\includegraphics[width=\columnwidth]{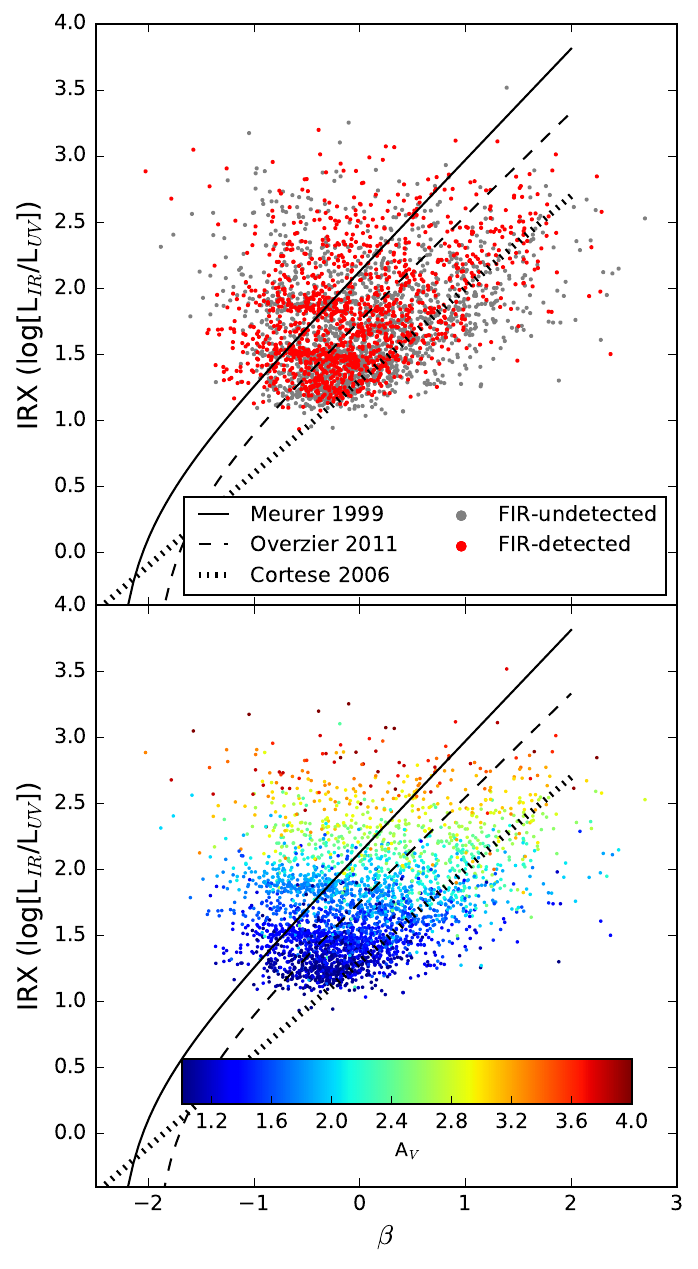}
\caption[width=\columnwidth]{Logarithm of IR to UV luminosity ratio as a function of UV continuum slope $\beta$, for our UVJ star-forming subsample. Several relations from the literature are shown as labeled. Top: \textit{Herschel} detections are shown in red and nondetections in gray. Bottom: Coloring indicates A$_V$. \label{fig1}}
\end{figure}

\begin{figure*}[ht]
\includegraphics[width=\textwidth]{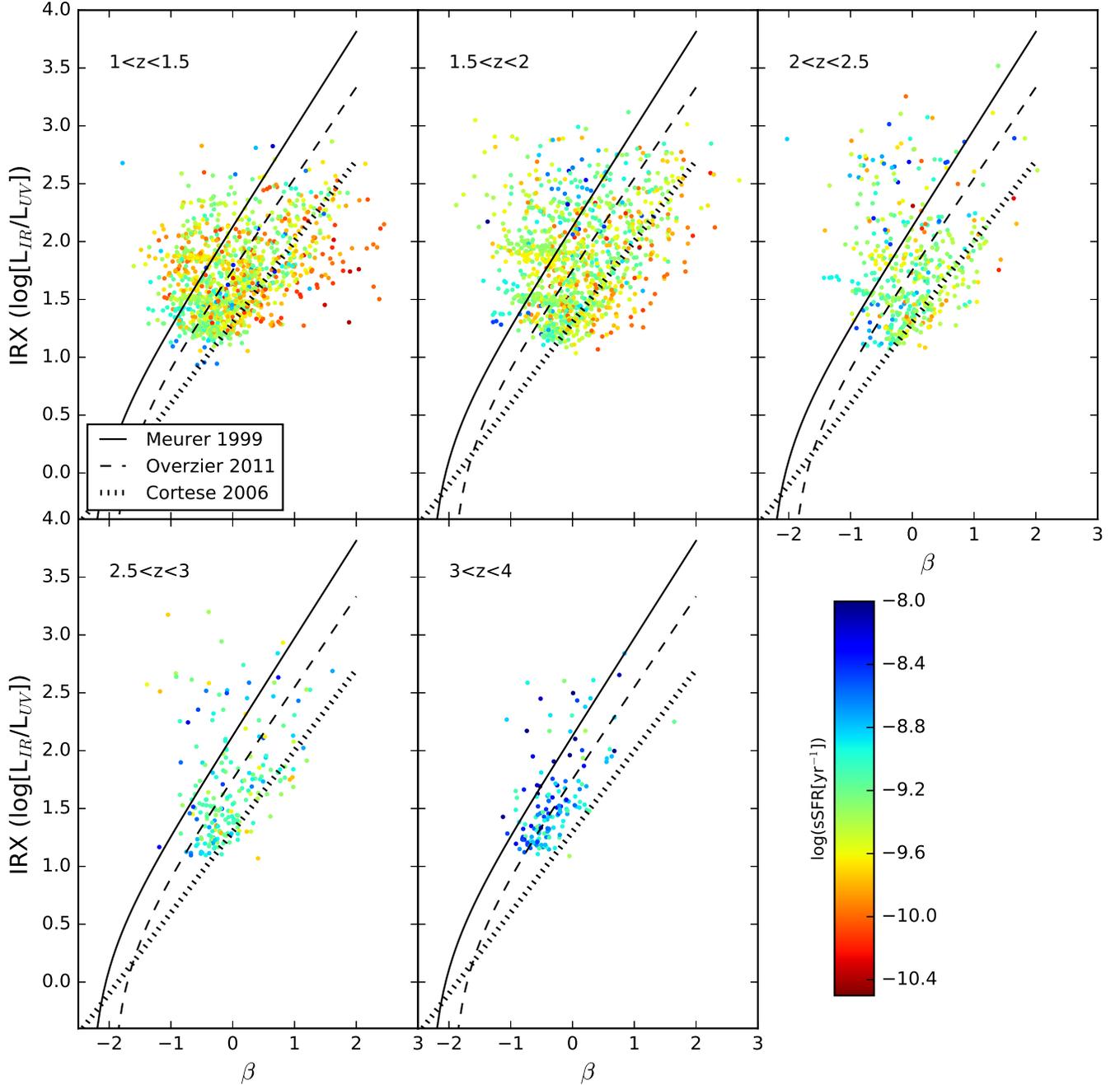}
\caption[width=\textwidth]{Logarithm of IR to UV luminosity ratio as a function of UV continuum slope $\beta$, for our UVJ star-forming subsample split into bins of redshift. Several relations from the literature are shown as labeled. Coloring indicates sSFR. \label{fig1}}
\end{figure*}

Given this correlation between dust luminosity and SFR, it is reasonable to expect the dust mass to scale with SFR as well.  In the top panel of Figure 10 we show this relation for our data with the additional parameter of the dust temperature indicated by color. This dust temperature is a weighted average of the temperatures of the two components of the dust model, namely the ISM and birth clouds. Using MAGPHYS, \citet{dacunha10a} found a tight correlation between dust mass and SFR in a local sample derived from the Sloan Digital Sky Survey (SDSS) which we show as a black line. The \citet{dacunha10a} sample is selected to be star-forming by emission line diagnostics and lies at $z \le 0.2$. It is also important to note that the SFRs for their sample lie mostly below $\sim 20$M$_\odot$yr$^{-1}$. The solid portion of the line represents the extent of their data. We also show the \textit{Spitzer} Infrared Nearby Galaxy Sample \citep[SINGS,][]{kennicutt03} analyzed in \citet{dacunha08} as purple stars for comparison. At first, our data do not appear to follow any trend, with only our highest measured dust masses falling near the \citet{dacunha10a} relation. Given that MAGPHYS allows a range of dust temperatures when fitting for the dust mass, we investigate the effects of dust temperature in this diagram. When we examine the temperature of the dust (which MAGPHYS allows to vary from $20-80 K$) as indicated by color in the figure, we see that sources with lower dust temperature actually lie on the published relation. If we consider bands of constant dust temperature in the diagram we see that the slope of the M$_{dust}$-SFR relation for our sources actually closely matches the slope of the \citet{dacunha10a} relation. As shown in Figures 6 and 7, we observe a slight increase in average dust temperature with redshift for our sample, so it may be tempting to interpret the variation in this diagram as a redshift dependence. We checked the relation between dust temperature and redshift for our sample and found that the scatter in temperature at a given redshift to be much larger than the evolution between redshift bins. This suggests that the local calibration between dust mass and SFR can be extended to higher redshift samples through the incorporation of the additional parameter of dust temperature. Given our particular sample selection, the application of these results to a more general star-forming population would have to be done with caution.

One may worry that the dust mass and dust temperature are degenerate if the FIR SED is not well sampled. To test whether the observed trend with temperature is driven by a lack of FIR detections, we show the same comparison in the bottom panel of Figure 10, now only showing sources detected in the FIR. As expected, we lose the low SFR tail in the diagram, but the trend with dust temperature is largely preserved. This suggests that the potential model degeneracy between dust mass and temperature in the case of sparse FIR sampling can be excluded as the only cause of the observed trend. We note that our median SEDs show some evidence for the evolution of dust temperature with redshift, but when we investigate the M$_{dust}$-SFR relation as a function of redshift the trend is much weaker than with dust temperature.

Previous work has shown a galaxy's sSFR to be a good tracer of several ISM properties in the local universe \citep{dacunha10a}. Here we test whether these relations hold at higher redshift. Figure 11 shows three ISM properties as functions of sSFR with color indicating the average dust temperature of each source. Gray contours show the distributions for our full sample while points show \textit{Herschel} detections. The top panel shows the dust to stellar mass ratio, the middle shows dust mass divided by SFR, and the bottom shows the fraction of infrared luminosity contributed by dust in the diffuse ISM ($f_\mu$). The distribution of the full sample from \citet{dacunha10a} is shown with the black contour, whereas their subsample of massive (log(M$_*$/M$_\odot$)$>10.6$) galaxies is shown with gray circles. We, observe a weak correlation between dust to stellar mass ratio and sSFR and anti-correlations in the other two cases. Our sample does not cover the low sSFR end of the distribution from \citet{dacunha10a}, although it does overlap in parameter space with their massive subsample. We observe a wider range in both dust to stellar mass ratio and dust mass divided by SFR due to sources with lower values in both cases. As in Figure 10, we see that the sources that fall below the distributions for the \citet{dacunha10a} sample are those with higher dust temperatures. The average dust temperature correlates with $f_\mu$ by definition since the allowed temperatures for birth clouds in the models are higher than those for the diffuse ISM. We find that our sample roughly matches the anti-correlation between $f_\mu$ and sSFR found in \citet{dacunha10a}. In all three cases the high redshift sources appear to continue the trends to higher sSFR.

\subsection{IRX-$\beta$ Relation}

One common measure of the dust properties of galaxies employed at high redshift is the so-called IRX-$\beta$ relation, where IRX is the IR to UV luminosity ratio and $\beta$ is the slope of the UV continuum such that $f_\lambda \propto \lambda^\beta$. Following \citet{whitaker14}, we estimate the total integrated 1216–3000$\AA$ UV luminosity by using the 2800$\AA$ rest-frame luminosity plus an additional factor of 1.5 to account for the UV spectral shape with the simplification of a 100 Myr old population with a constant SFR, where $L_{\rm UV}(1216–3000 \AA) = 1.5\nu L_{\nu,2800}$. The rest-frame luminosity at 2800$\AA$ is calculated from the best-fit EAZY SED via the method described in \citet{brammer11}. We calculate $\beta$ by fitting a power law to the rest-frame fluxes at 1400, 1700, 2200, 2700, and 2800 $\AA$ in the SED fits from EAZY. The top panel of Figure 12 shows the IRX-$\beta$ relation for only our UVJ star-forming galaxies with FIR-detected galaxies in red and nondetections in gray. The canonical relation for starbursts from \citet{meurer99}, its update in \citet{overzier11},  and the corresponding relation for normal star-forming galaxies from \citet{cortese06} are shown for reference. The lack of a FIR detection does not appear to bias the location of the sample in parameter space for our selection. Our data show a large dispersion around the local relations, and in fact no strong correlation between IRX and $\beta$ is observed. In order to compare position in the diagram with other measurements of extinction, the bottom panel of Figure 12 shows the A$_V$ determined by MAGPHYS for each source. We see a clear trend of increasing A$_V$ with increasing IRX. Interestingly, it shows no dependence on $\beta$. We verified that the same behavior is observed for A$_{1600}$ when it is calculated in the same manner as A$_V$. 

The evolving locations of sources in our UVJ diagrams (Figures 2,3) suggest a possible evolution in dust properties with redshift. To test this idea, we show in Figure 13 the IRX-$\beta$ relation for UVJ star-forming sources in bins of redshift with sSFR indicated by color. First, we observe as before an elevation of sSFRs with redshift. Up to $z \sim 2.5$ we see the same wide scatter in the diagram, although at fixed redshift the most vigorously star-forming sources tend lie on the left of the diagram with low $\beta$ values, and conversely for the least star-forming sources. At higher redshift, the spread in $\beta$ shrinks considerably, such that for our highest redshift bin our sample now appears to be consistent with the \citet{meurer99} relation. Given that the sources in this bin have $log(sSFR[yr^{-1}]) \gtrsim -9$ they may more closely resemble the starburst sample on which the relation is based.

\section{Discussion}
\subsection{Dust Emission in Quiescent Galaxies}

\begin{figure*}[ht]
\includegraphics[width=\textwidth]{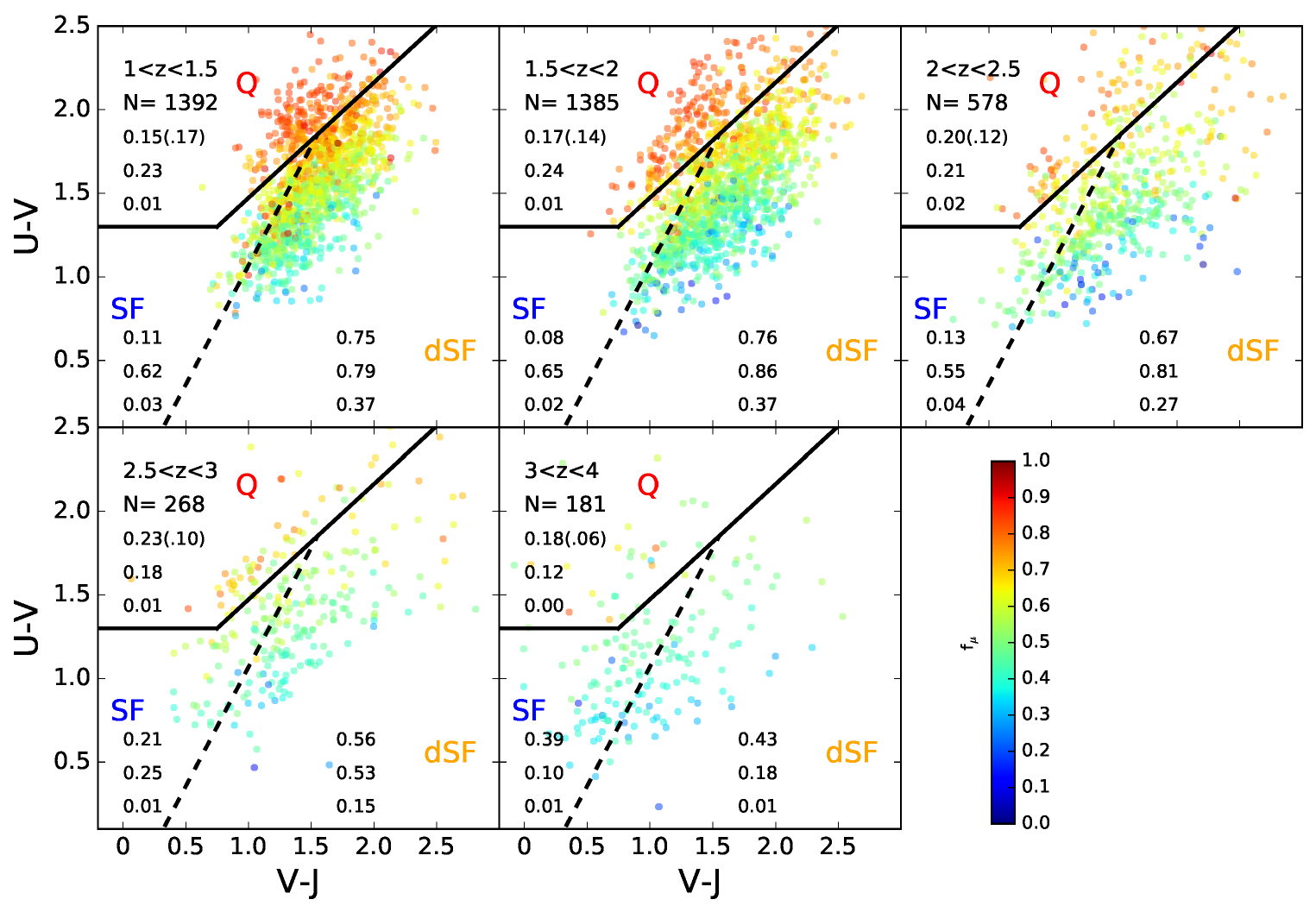}
\caption[width=\textwidth]{UVJ diagram split by bins of redshift as in Figures 2,3. Coloring indicates f$_\mu$. \label{fig1}}
\end{figure*}

The difference between star-forming and quiescent galaxies in the L$_{\rm IR}$-SFR relation shown in Figure 9 suggests that caution should be used when interpreting the infrared emission of massive galaxies at high redshift since intermediate age to evolved stars can significantly contribute to the observed IR emission. Indeed, \citet{utomo14} find that simply summing UV and IR SFR indicators leads to an overestimation of the SFR compared to simultaneous modelling of stellar and dust emission.  \citet{fumagalli14} previously showed through a stacking analysis of MIPS $24\micron$ data that emission of circumstellar dust envelopes and cirrus dust heated by evolved stars could account for the excess $24\micron$ emission with regard to their measured SFRs. They also note that this effect would increase with increasing redshift due to the combination of younger stellar ages and increasing A$_V$ \citep[see also, the increasing fraction of dusty star-forming galaxies in][]{martis16}. Still, even though our UVJ quiescent galaxies have IR emission dominated by the ISM, they typically possess SFRs of the order of a few $M_\odot$ yr$^{-1}$ (corresponding to sSFR $\sim 10^{-10}$ yr$^{-1}$). While this is about an order of magnitude below the typical values for the star-forming galaxies in this sample, it demonstrates that dusty UVJ quiescent sources can still host residual star formation at high redshift. Similarly, \citet{fumagalli14} also find that when stacking dusty star-forming SEDs with quiescent SEDs, the resulting colors can produce measurements that remain in the quiescent box of the UVJ until the contribution of the dusty star-forming component reaches as high as 30\%. 

It should also be noted that the SFR calculated here by MAGPHYS is averaged over the past 10 Myr, so sources identified as quiescent here with relatively low sSFRs may have had episodes of star formation in the recent past. If this is the case, some of our quiescent sources may be better described as post-starburst galaxies and their substantial dust content could be the result of recent star formation. Simulation and observational studies have shown L$_{\rm IR}$ to overestimate the instantaneous SFR particularly severely for galaxies with declining star-formation histories \citep{hayward14, sklias17}, but that MAGPHYS can reliably recover physical parameters for both isolated disk galaxies and mergers provided the modeling assumptions (e.g., dust law) do not radically differ from the simulation input \citep{hayward15}.

To more directly investigate both the origin and detectability of MIR emission for our sample, Figure 14 shows the UVJ diagram for our sample as in Figures 2 and 3. Coloring now indicates f$_\mu$. We remind the reader that the numbers in each panel show the fraction of sources in each region and the fractions within that region detected at 24 $\micron$ and by \textit{Herschel} respectively. First, we see a clear gradient in f$_\mu$ across the diagram. This shows that the dust luminosity in the star-forming region is dominated by the birth cloud component of the dust model, whereas dust emission in the quiescent region is dominated the by ISM heated by evolved stars. As expected, the incidence of MIPS detections is higher for star-forming sources at all redshifts. Nevertheless, we see significant detection fractions for quiescent sources as well, reaching as high as 32\% in our $1.5 \le z < 2$ bin. As the coloring shows, the MIR emission from these sources is dominated by dust heated by evolved stars, mostly over 80\% for our two lowest redshift bins. Consequently, a direct conversion of a 24 $\micron$ flux into a SFR for these sources would clearly overestimate the level of star-formation. This figure shows that the interpretation of IR emission from massive, dusty galaxies at high redshift must be done with care and that simultaneous modeling of the panchromatic SED is one effective method to overcome this difficulty. 

The degree to which panchromatic SED modeling allows for a corrected conversion of L$_{\rm IR}$ to SFR as in the middle panel of Figure 9 depends on the accuracy with which the source of radiation can be determined. Our ability to bring star-forming and quiescent galaxies onto the same relation indicates that the $f_\mu$ parameter from MAGPHYS accomplishes this fairly robustly for samples of galaxies, but should probably be used with caution for individual sources. The median error on $f_\mu$ estimated by MAGPHYS is $\sim 0.13$ for our sample regardless of far-IR detection. The fairly wide scatter of the distribution of star-forming sources in the L$_{\rm IR}$-SFR plane in the middle panel of Figure 9 may be the result of uncertainty in $f_\mu$.

\subsection{Evolution of Dusty Galaxies}

Figure 11 shows that the correlations between sSFR and ISM properties are extended at the high sSFR end by high-redshift galaxies (Figure 2 shows that most of our sources with the highest sSFRs lie in our upper redshift bins). Since this increasing sSFR with redshift corresponds qualitatively with the observed evolution of the star-forming main sequence, it is possible to suggest an evolutionary explanation for these relations. This requires the assumption that our massive and dusty selection corresponds to populations directly connected by evolution. With this caveat, we can expect that as they age, galaxies would move leftward along the trends in these plots. When considering the dust to stellar mass ratio, this would imply that intense star formation activity is associated with the production of large amounts of dust. As a galaxy's stellar mass is built up and star formation slows down, the dust to stellar mass ratio will decrease. From this plot we cannot determine the degree to which destruction of the dust associated with the starburst, or creation of more dust by evolved stars, would affect this relation. From the coloring in the plot, we see that the average dust temperature scales inversely with the dust to stellar mass ratio, with our sources spanning $\sim 2$ dex in this quantity. 

Given the relation of the SFR density to gas density via the Kennicutt-Schmidt law \citep{kennicutt98}, \citet{dacunha10a} interpret the dust mass divided by SFR as a proxy for a dust to gas ratio. In this case, a galaxy moving along the trend in the middle panel of Figure 11 would build up its dust mass as it depletes its gas reservoir. This scenario is consistent with the interpretation of the top panel. If the SFR does indeed trace the gas density in our sample, then we see that a wide range of dust to gas ratios are possible. Potential explanations for this wide span could be due to metallicity or initial mass function variations, but we lack the data to explore this question further. Given our results in Figure 10, the trend of decreasing dust temperatures with increasing dust mass to SFR ratios is unsurprising. One possible explanation for this trend is dust destruction at high temperatures. This effect would depend on the species of dust grains and so remains an interesting avenue open to further research. 

The bottom panel of Figure 11 supports our interpretation of Figure 9 that $f_\mu$ traces the amount of star formation. We see that higher values of $f_\mu$ correlate with lower sSFR, or that for galaxies in which star formation is shutting down, more of the dust luminosity comes from the diffuse ISM heated by intermediate age to evolved stars rather than birth clouds. Additionally, from Figure 10 we know that the dust temperature also factors into the relation between the dust mass and SFR. One possible interpretation for this trend is shielding of the ISM dust by the birth clouds. As the amount of dust builds up, less radiation from stellar birth clouds will be able to leak out, resulting in a weaker radiation field in the ISM and correspondingly lower temperatures for a given SFR. At a fixed dust mass, the dust temperature does appear to correlate with the SFR, but a significant temperature gradient corresponds to changes in dust mass. 

Although we see a trend toward increasing dust to stellar mass ratios with increasing sSFR, we observe high dust masses even for galaxies with low sSFR (Figures 10, 11). In a recent study of massive, quiescent galaxies in COSMOS, \citet{gobat18} also observe much higher dust to stellar mass ratios ($M_{dust}/M_* \sim 8 \times 10^{-4}$ at $z = 1.76$) than reported for local early type galaxies \citep[$M_{dust}/M_* \sim 10^{-5}-10^{-6}$,][]{smith12}. The values we observe  of $M_{dust}/M_* \sim 10^{-3.5}$, at the low end of our sSFR distribution agree well with those reported by \citet{gobat18}, even though our samples differ in that it is pre-selected to be dusty, with A$_V>1$ mag. We also note that our dust to stellar mass ratios for sources with lower dust temperatures agree well with those from the massive subsample of \citet{dacunha10a} which is a local sample, so such sources do not seem to be confined to high redshift. 

Several previous studies have identified a trend of increasing dust temperature with redshift for star-forming galaxies \citep{magdis12, bethermin15, schreiber17}. In Figure 6 we do not observe a strong shift in the peak of the FIR continuum to shorter wavelengths with increasing redshift at $z<2.5$ in the median SEDs, although the scatter in SED shape makes this an uncertain result. Moreover, we note that Figure 6 includes all massive and dusty galaxies, i.e., both star-forming and quiescent galaxies as identified by the UVJ diagram, with the fraction of quiescent galaxies in the range 6-23\% depending on redshift and on the specific definition of quiescent galaxies (i.e., using the UVJ diagram or the sSFR). The inclusion of quiescent galaxies may dilute the detectability of the dust temperature increase with increasing redshift. In fact, we see in Figure 7 that UVJ star-forming galaxies have IR peaks at shorter wavelengths than their quiescent counterparts, indicating that at a given redshift active star formation corresponds to a higher dust temperature. Even so, comparing the SEDs of only UVJ star-forming galaxies in Figure 7 at different redshifts still shows only weak evidence for an evolution in dust temperature at $1<z<2.5$, since the far-IR SEDs for the massive and dusty star-forming galaxies peak at 72$\mu$m, 71$\mu$m, and 66$\mu$m at $1.0<z<1.5$, $1.5<z<2.0$, and $2.0<z<2.5$, respectively. The median SEDs for the two highest redshift bins peak at significantly shorter wavelengths of $\sim 50\mu$m and $30 \micron$, respectively. \citet{bethermin15} obtained similar results for a sample of massive star-forming galaxies, observing a shift in the far-IR peak from $\sim 70 \micron$ at $z \sim 1.1$ to $\sim 30 \micron$ at $z \sim 3.75$. Additionally, \citet{bethermin15} also reported a broadening of the peak of the dust emission at $z > 2$, in agreement with our findings at $2.5<z<4$. These shifts in peak wavelength in the far-IR SED are consistent with the increase of the median weighted dust temperature of the massive and dusty star-forming galaxies from $39 K$ to $43 K$ over the entire redshift range $1.0<z<4.0$. We note however that \citet{schreiber17} finds cooler temperatures and stronger redshift evolution (from $27 K$ to $39 K$, over a similar redshift range and stellar mass regime) than us, although a direct comparison is difficult to make given the differences in the galaxy samples.

Lastly, Figure 14 shows that in addition to correlating with position in the UVJ diagram, our distribution of $f_\mu$ values shifts higher toward lower redshift. Additionally Figure 11 shows that this increase in $f_\mu$ corresponds to a decrease in sSFR. 

\subsection{Are FIR-selected Samples a Unique Population?}

We remind the reader that 'dusty' as defined in this work corresponds to $A_V \ge 1$, a property of the rest-frame optical region of the SED, so it is not clear a priori that this selection is commensurate with a selection using the FIR. Figure 4 shows that for the parameters which we investigate, FIR-selected sources do not occupy a different region of parameter space from the general dusty galaxy population. Rather these sources make up the upper ends of continuous distributions in stellar mass, SFR, dust extinction, and dust mass. This agrees with previous observations of increasing obscuration with stellar mass \citep{whitaker12, martis16}.

From Figures 9, 10, and 12 we observe no discernible difference between our \textit{Herschel}-detected sources and the rest of our sample in terms of the L$_{\rm IR}$-SFR, M$_{dust}$-SFR, and IRX-$\beta$ relations we investigate. This similarity in ISM properties would seem to be evidence against a differing mode of star formation for our FIR-detected subsample. At face value this seems to disagree with suggestions of unique ISM conditions in luminous starbursts at $z \ge 1$ as has been suggested in the literature \citep[e.g.][]{bethermin15}, but we actually have relatively few sources that would qualify as strong starbursts for our observed redshift range (for comparison, the \citet{bethermin15} starburst sample has sSFR $\sim 10^{-8}$yr$^{-1}$. A cut in SFR would lead to an overlapping, but different selection of sources from our FIR cut (Fig. 4). The selection of our sample as both massive and dusty obviously influences this comparison as well. It is possible that if we extended our investigation to less extreme objects, that a substantial difference between the majority of objects and those detected by \textit{Herschel} could be seen. Our current results, however, suggest that \textit{Herschel}-detected sources do not differ qualitatively in their ISM properties from the general massive, dusty population at high redshift.  

Other than increased stellar mass and SFR, our work cannot constrain other possible parameters affecting dust attenuation or emission properties. The dependence on stellar mass may correspond to a correlation of dust properties with metallicity via the stellar mass-metallicity relation. But actually, under the na\"ive assumption that dust mass traces metallicity, the reduction of scatter in the mass-metallicity relation through the incorporation of SFR \citep{mannucci10} seems at odds with our observed correlation between SFR and dust mass. This may be explained by the formation and dissipation of stellar birth clouds without enrichment of the surrounding ISM which could be accomplished later by more evolved stars. Verification of any such process as well as a determination of the relationship between dust and metallicity would require spectroscopy which remains costly particularly for heavily obscured sources. One might also expect the increasing compactness of star-forming galaxies with redshift \citep{vanderwel14} to lead to higher attenuation levels. A related geometric explanation for obscuration levels may be related to the concentration of star-forming gas during major mergers. 

\subsection{Accounting for Dust in High-z Studies}

Investigation into both the form and physical origin of the IRX-$\beta$ relation has been an active area of research in recent years. The relation is particularly important for studies of star formation at high redshift since it is frequently used to correct observed UV-based SFRs for dust obscuration. In our sample, which should be recalled to be dustier than typical Lyman break galaxies which are commonly used for estimating the cosmic SFR density at $z > 5$, we find important differences from the local reference relations. Our dusty selection is biased toward high IRX and $\beta$ values so that we do not sample the full range commonly explored for these parameters. Even the NIR-selected sample of \citet{reddy18} has an average $\beta \sim -1.7$ for their full stack, which lies at the low edge of our distribution of $\beta$ values. Additionally we observe a large dispersion around the local relations such that our sample actually does not appear to follow any clear relation between IRX and $\beta$ when we consider sources up to $z \sim 3$. 

If we compare our sample to others from the literature, we find that our wide range of observed $\beta$ values is consistent with the IR-selected sample of dusty star-forming galaxies from \citet{casey14}, but our requirement of $A_V \ge 1$ limits our minimum observed IRX to higher values. \citet{casey14} find their sample of dusty star-forming galaxies to be shifted to bluer $\beta$ than their local comparison sample, in line with the IR-selected samples of \citet{reddy12} and  \citet{penner12}. Our NIR-selected sample has a slightly lower typical IRX but similar $\beta$. In the overlapping redshift range, \citet{casey14} do not observe significant redshift evolution, which agrees qualitatively with the main locus of our sources remaining at $\beta \sim -0.5$ in all of our redshift bins. For the limited region of overlap in IRX values between the \citet{reddy18} NIR-selected sample and ours, we observe a similar typical $\beta \sim -0.4$.

These findings are obviously dependent on the assumptions we have made regarding the intrinsic UV slope through our choice of stellar population synthesis model \citep{bc03} and attenuation law \citep{charlot00}. The star formation history, stellar mass, and metallicity which depend on the assumed stellar population synthesis model will change the intrinsic UV slope and hence the observed IRX-$\beta$ relation \citep{reddy12, grasha13, schaerer13, reddy18}. Additionally, observational evidence has suggested that dust obscuration can be decoupled from the UV slope in the case of large SFRs \citep{penner12, casey14} and that the spatial configuration of the stars and dust can lead to a wide range in $\beta$ \citep{koprowski16}. A comprehensive analysis of the IRX-$\beta$ relation and its uncertainties for our sample is beyond the scope of this work; rather we wish to provide a point of contrast to the commonly studied UV-selected samples at high redshift.

Figure 12 indicates that A$_V$ correlates with IRX, but not $\beta$. This is consistent with \citet{narayanan18} who in a study of cosmological zoom-in simulations find IRX to be a better tracer of UV optical depth than $\beta$. They also identify complex dust geometries and aging stellar populations as drivers of scatter in the IRX-$\beta$ relation, both of which likely affect our sample of very dusty and massive galaxies. This is an important result for the interpretation of corrected UV star-formation rates at high z.

Finally, we note the improved agreement between our sample and the local IRX-$\beta$ relation for starbursts in our highest redshift bin. We have observed that the sources in this bin have higher specific star formations rates, bluer U-V and V-J colors, and lower $f_\mu$ values than the majority of our sample. From our median SED for this bin in Figure 6, we see evidence for a strong break near Ly-$\alpha$. Given this and their blue colors, these sources may be more similar to the Lyman break galaxies that are used to infer the cosmic SFR density at $z \ge 4$ \citep[e.g.][]{steidel99, bouwens09, castellano12}. If this is the case, then their agreement with established dust correction relations despite their apparent significant obscuration bodes well for studies of UV-selected galaxies at high redshift, with the caveat that our sample has been selected very differently. 

\begin{figure*}[ht]
\includegraphics[width=\textwidth]{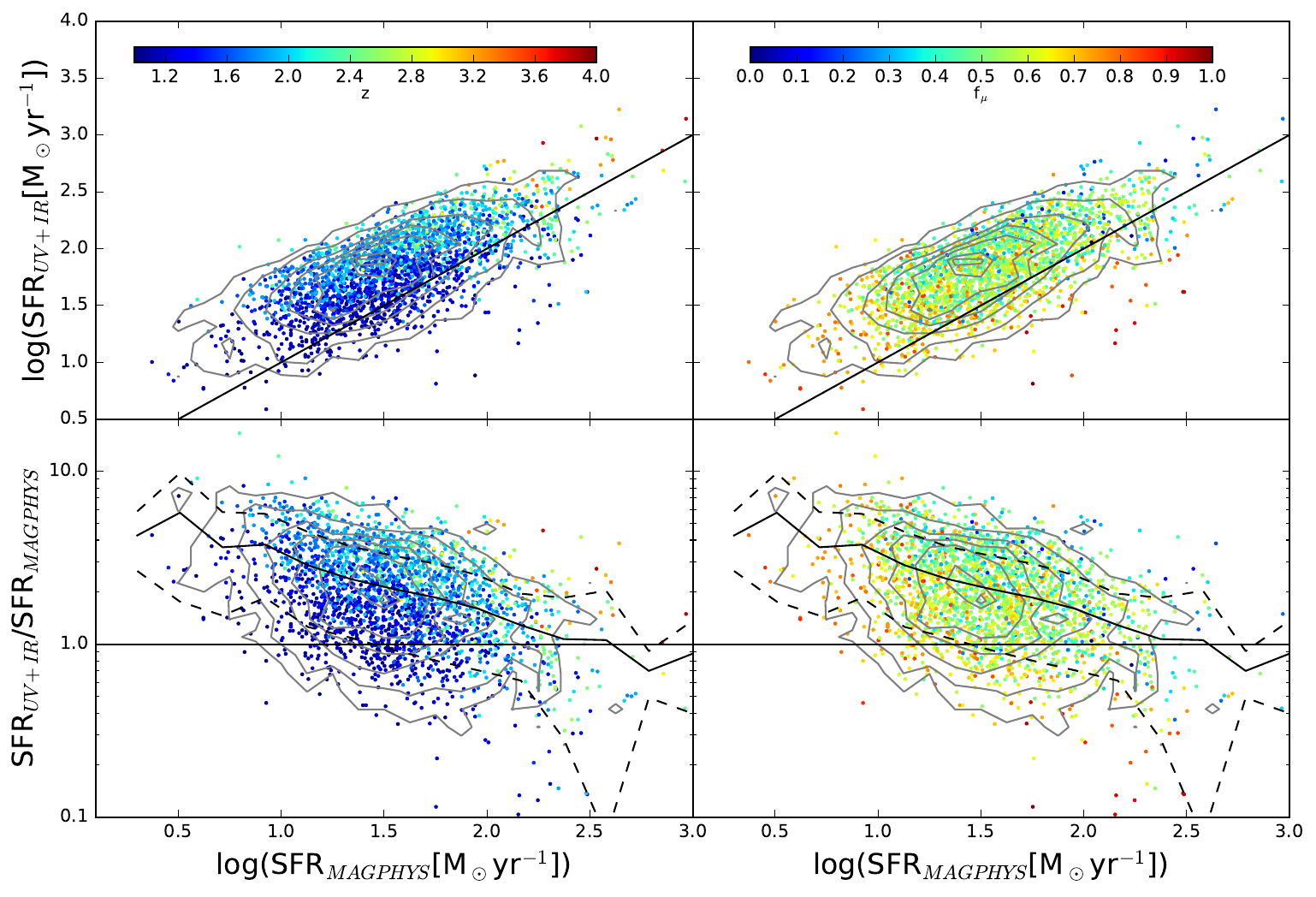}
\caption[width=\textwidth]{Top: UV+IR SFR where IR luminosity is calculated from the MIPS 24$\micron$ flux and the \citet{dale02} templates vs SFR from MAGPHYS. Contours show the distribution for all MIPS-detected sources while points show those that are classified as star-forming by UVJ colors only. Contours are placed at the 5$^{th}$, 15$^{th}$, 35$^{th}$, 50$^{th}$, 65$^{th}$, 85$^{th}$, and 95$^{th}$ percentile levels. Coloring in the left and right columns indicate redshift and f$_\mu$ respectively. Bottom: As above, but now showing the ratio of UV+IR and MAGPHYS SFRs. The running median and 1-$\sigma$ scatter are shown with solid and dashed black curves. \label{fig1}}
\end{figure*}

\subsection{SFR$_{MAGPHYS}$ vs. SFR$_{UV+MIPS}$}

We have seen that a substantial fraction of L$_{\rm IR}$ for massive, dusty galaxies arises from sources other than recent star formation (e.g. Fig. 9 and 14). Additionally we have shown that even within this class of galaxies, we observe a range of SED shapes depending on the level of star formation, redshift, and the contribution of evolved stars to dust heating (Fig. 6, 7, and 8). Both of these results highlight important differences within the overall galaxy population that are ignored when adopting the common practice of using a template to scale a single flux measurement to an IR SFR. To illustrate the danger of such a method as commonly implemented in the literature we show in Figure 15 a comparison of UV+IR SFR calculated in this manner and the SFR obtained from MAGPHYS. To calculate the UV+IR SFR, we follow the prescription from \citet{bell05} used in \citet{whitaker14} so that 
\begin{equation}
    SFR[M_\odot yr^{-1}] = 1.09 \times 10^{-10}(L_{\rm IR}[L_\odot] + 2.2L_{\rm UV}[L_\odot])
\end{equation}
where L$_{\rm IR}$ is calculated from the MIPS 24$\micron$ flux calibrated by the \citet{dale02} templates and L$_{\rm UV}$ is as determined above for the IRX-$\beta$ relation. Specifically, following \citet{marchesini10} and \citet{whitaker10}, we computed the total infrared luminosity for each object for all \citet{dale02} templates; the mean of the resulting log(L$_{\rm IR}$) was adopted as a best estimate for L$_{\rm IR}$. Systematic template uncertainty can be as high 0.4-0.5 dex due to the range of the \citet{dale02} templates \citep{wuyts08}. The top row of Figure 15 shows the UV+IR SFR versus MAGPHYS SFR for all MIPS-detected sources with contours. UVJ star-forming MIPS sources (which make up the majority of MIPS sources) are shown as colored points. Coloring in the left column indicates redshift, and in the right column indicates f$_\mu$, the fraction of IR luminosity originating from the diffuse ISM. The UV+IR SFR clearly overestimates the level of star formation for most sources. This overestimation is due to the IR portion of the SFR since the typical contribution of the UV SFR is only $\sim 5\%$ for our sample of dusty galaxies. This is also evident from Figure 12 which shows that the minimum IRX for our sample is $\sim 10$.

In order to better quantify the difference between the two SFR measurements, we show in the bottom row of Figure 15 the ratio of UV+MIPS and MAGPHYS SFRs as a function of MAGPHYS SFR. Again all MIPS sources are shown with contours and UVJ star-forming sources with points matching the coloring in the panels above. The running median and 1-$\sigma$ scatter are shown as solid and dashed black curves. We see that the disagreement between the two estimators is actually a strong function of SFR. This can be understood in terms of f$_\mu$. At low SFRs, f$_\mu$ is high, meaning that a large fraction of L$_{\rm IR}$ originates from the diffuse ISM. Converting the total observed L$_{\rm IR}$ into a SFR then leads to an overestimation by a factor of $\sim 4$. For the majority of our sample, we observe intermediate f$_\mu$ and SFR values on the order of tens M$_\odot$yr$^{-1}$, leading to a less severe, but still substantial overestimation of a factor of $\sim 2-3$ from the UV+MIPS method. At the highest SFRs, which correspond to the highest redshift sources, the majority of L$_{\rm IR}$ is actually due to star formation, making the direct conversion of 24$\micron$ flux to a SFR a valid approximation statistically. We note however, that for $SFR_{MAGPHYS} \gtrsim 100 M_\odot yr^{-1}$ there is evidence of a different behavior for our high- and low-redshift objects. Specifically, the UV+MIPS SFR overestimates the SFR from MAGPHYS for high redshift objects and underestimates for low redshift sources. This further illustrates that the assumption of a single FIR template over a wide redshift range may not generally hold for massive dusty galaxies. We remind that template uncertainty in the adopted method to estimate SFR from MIPS 24 micron fluxes can introduce systematic uncertainties as large as 0.4-0.5 dex.. 

A few caveats apply to this comparison. First, the UV+MIPS SFR only uses the MIPS flux for L$_{\rm IR}$, but the MAGPHYS modeling includes \textit{Herschel} data when available, and the upper limits otherwise, so the estimates of L$_{\rm IR}$ use different regions of the observed SED. Also, Figure 1 shows that the distribution of f$_\mu$ values for our massive, dusty galaxies is strongly offset from the majority of the NIR-selected parent sample which better reflects the majority of samples to which this method of UV+IR SFR estimation is usually applied. The lower f$_\mu$ values of the parent sample should result in a less severe disagreement between the SFR estimations. This reinforces the need to be wary of the effects of sample selection when applying relations from the literature to other bodies of work.

Additionally, using the new Prospector-$\alpha$ SED modeling software, \citet{leja19} find similar results when applying their code to $\sim 58,000$ galaxies at $0.5 \le z \le 2.5$ from the 3D-HST survey. Their panchromatic SED modeling finds sSFRs systematically lower than those implied by UV+IR estimates, with the strongest offset reaching up to 1 dex at the lowest sSFRs. Although multiple factors contribute to the offset, they identify emission from stars older than 100 Myr as the primary cause. At lower sSFR old stars contribute a higher fraction of the observed luminosity, and thus UV+IR becomes an increasingly worse estimate of the SFR (see their Figures 7 and 8). This is consistent with our finding of the largest offset between SFR$_{MAGPHYS}$ and SFR$_{UV+MIPS}$ at low SFR as well as our observed correlation of f$_\mu$ with the sSFR. \citet{boquien16} also find the calibration of UV+IR SFR estimators to strongly vary with both stellar mass density and sSFR due to dust heating by older stellar populations. In a simulation study of both isolated disks and mergers Roebuck et al. (accepted) find a color-based SFR correction based on the contribution of older stars on a per object basis to much more accurately recover the true values than fixed scaling relations of luminosity to SFR. Thus not only do older stars appear to be important to the estimation of physical properties for our particular sample, there is a mounting body of evidence that they should be accounted for in a more nuanced approach in more general studies than is currently typically done.

\section{Summary}

We combined the UltraVISTA DR3 multi-wavelength UV-to-MIPS 24 $\micron$ photometric catalog (Muzzin et al. in prep.) with \textit{Herschel} PACS-SPIRE photometry \citep{oliver12, lutz14, hurley16}, and we modeled the resulting UV-to-FIR observed SEDs using MAGPHYS \citep{dacunha08} to investigate the stellar population and dust properties of a complete sample of massive (log(M$_*$/M$_\odot) \ge 10.5$) and dusty (A$_V \ge 1$ mag) galaxies at $1.0 \le z \le 4.0$, and to explore their relation to FIR-selected samples of galaxies. We find the following main results.

\begin{enumerate}
  \item The UVJ diagram is found to trace reasonably well the star-formation rates and Av (as derived by MAGPHYS) out to $z<3$, with massive dusty galaxies mostly populating the dusty star-forming region in the UVJ diagram ($\sim 60-80\%$) and $\sim$15-20\% of the sample being classified as quiescent galaxies in the UVJ diagram (in agreement with their sSFRs and comparatively suppressed SFRs). The UVJ-selected quiescent galaxies appear to have significant dust obscuration ($1<$A$_V[mag] \le 2$). Although we cannot exclude some level of (expected and unavoidable) contamination from dusty star-forming galaxies, our findings are consistent with recent results from Gobat et al. (2018). Differently from $z<3$, at $3<z<4$, our sample of massive dusty galaxies preferentially ($\sim$50\%) populate the region in the UVJ diagram of unobscured star-forming galaxies, with $\sim$40\% in the dusty star-forming region. This could be an indication that the UVJ diagram should be used with care at $z>3$, where massive galaxies may be more likely to have complex dust geometry, potentially with spatial separation of the regions generating the rest-frame optical-NIR emission (and sampling the rest-frame UVJ) and the IR emission (constraining the SFR).
  
  \item FIR-selected samples occupy the upper end of continuous distributions in SFR, stellar mass, and dust luminosity compared to the general massive, dusty population.
  
  \item The median SED of massive, dusty galaxies evolves strongly with redshift, toward weaker MIR and UV emission as well as redder UV slopes with increasing cosmic time.
  
  \item The relative dust luminosity contributed by the ISM compared to stellar birth clouds (f$_\mu$) strongly correlates with SED shape for massive, dusty galaxies. Specifically, larger values of $f_\mu$ (i.e., less stellar birth cloud contribution to the IR emission) corresponds to smaller rest-frame UV emission and redder UV slopes, weaker MIR features, decreasing dust continuum emission, and peak of the IR emission shifting to longer wavelengths, suggesting lower dust temperatures. This is consistent with $f_\mu$ anti-correlating with the star-formation rate.
  
  \item For most objects in our sample of massive and dusty galaxies at $1<z<4$, one cannot blindly apply the \citet{kennicutt12} relation between L$_{\rm IR}$ and SFR. Good agreement with Kennicutt's (2012) relation is found only if the contribution to the IR emission from birth clouds is considered alone, after subtracting out the contribution from the ISM. Our findings indicate that in most massive and dusty galaxies there is a significant amount of dust emission not connected directly to star formation, but caused by dust heating from a more evolved stellar population. Therefore, the interpretation of the IR emission from dusty massive galaxies at high redshift must be done with care, and simultaneous modeling of the panchromatic SED is one effective method to overcome the complexity of these systems.
  
  \item The local relation between dust mass and SFR is followed by a sub-sample of massive and dusty galaxies with cooler dust temperatures, but not for objects with warmer dust temperatures, which show significantly reduced dust masses at a given SFR with respect to the local relation. Therefore, knowledge of the dust temperature is required to extend the local relation between dust mass and SFR to higher redshifts.
  
  \item The star-forming galaxies in our sample of massive and dusty galaxies at $1<z<3$ do not seem to follow the IRX-$\beta$ relation. However, we find better agreement at $3<z<4$ and that IRX strongly correlates with extinction (e.g., A$_V$ or A$_{1600})$.
  
  \item Our sample of massive dusty galaxies at $1<z<4$ follow local relations between dust-to-stellar mass ratio, ratio of dust mass and SFR (proxy for dust-to-gas ratio), and $f_\mu$ as a function of sSFR, albeit with large scatter that correlates with average dust temperature. Our high-redshift sub-sample extends these relations to the high sSFR end with respect to local values.
  
  \item The FIR-detected subset of our sample does not discernibly differ in ISM properties or scaling relations from the rest of our massive and dusty sample, suggesting no qualitative difference in the mode of star formation between these two populations. 
  
  \item Simple summation of UV and IR SFRs via the application of template scaling of L$_{\rm IR}$ leads to systematic overestimation of the SFR calculated from our panchromatic SED modeling. The overestimation is worst for our least star-forming objects, being a factor of $\sim 4$ higher. Statistical agreement is better for the highest SFRs in our sample, but still carries large uncertainties that appear to correlate with redshift.
\end{enumerate}

Spatially resolved UV-to-IR observations of these dusty and massive galaxies would be one way to address some of the shortcomings of the presented analysis and to further test some of the presented results. Spatially resolved \textit{ALMA} data (both continuum and spectroscopy) would be ideal to better characterize these objects and their ISM properties. Given the level of dust obscuration, \textit{ALMA} spectroscopy may be the most efficient way to obtain spectroscopic redshifts for this sample. Deep rest-frame optical spectroscopy (perhaps with \textit{JWST}) would be also needed to confirm the existence of this very intriguing population of high-z quiescent galaxies with significant dust obscuration in the UV-to-NIR photometry. 

\acknowledgments
D.M. and N.M. acknowledge the National Science Foundation under grant No. 1513473. 
This research has made use of data from HerMES project (http://hermes.sussex.ac.uk/). HerMES is a Herschel Key Programme utilising Guaranteed Time from the SPIRE instrument team, ESAC scientists and a mission scientist.
The HerMES data was accessed through the Herschel Database in Marseille (HeDaM - http://hedam.lam.fr) operated by CeSAM and hosted by the Laboratoire d'Astrophysique de Marseille.
Herschel is an ESA space observatory with science instruments provided by European-led Principal Investigator consortia and with important participation from NASA.
\\
Software: python, astropy, matplotlib, numpy, MAGPHYS, EAZY

%Facilities: \facility{VISTA}, \facility{Herschel}.


\appendix
\section{A: Effect of \textit{Herschel} Photometry on MAGPHYS Modeling}

The majority of the sources in our sample are not detected with \textit{Herschel}. In this Appendix, we attempt to quantify the amount of systematic effect introduced by the lack of \textit{Herschel} photometry. To accomplish this, we selected the 49 sources from our massive and dusty sample which are detected in all five \textit{Herschel} bands and reran MAGPHYS with those bands excluded. These sources have the most robust measurements of the FIR SED to allow us to observe the impact of the FIR data on the modeling. In Figure 16 we quantify the effects of these modeling differences on the MAGPHYS output parameters by directly comparing them. In each panel, the red points and curve show the running median values, while the 16$^{th}$ and 84$^{th}$ percentiles of the distribution are shown by the filled red regions. The average error of the points in each bin are shown by the red error bars, while the black error bars represents the average error of all plotted points, also specified by the written numbers. The median and standard deviation of the offset are indicated in the top left of each panel. We find that the stellar masses are robustly determined. For the IR luminosity and the SFR, we do find systematically larger values when the Herschel data are included, but the inferred offsets (0.16 dex and 0.19 dex for SFR and L$_{dust}$, respectively) tend to be smaller than the scatter (0.28 dex and 0.19 dex for SFR and L$_{dust}$, respectively). Especially for the SFR, the significance of the offset is only at the ~1.5 sigma, as the typical uncertainty is $\sim$0.1 dex. For A$_V$, we do find larger values when the Herschel data are included (by 0.35 mag), but this is also smaller than or comparable to the scatter (0.39 mag). For f$_\mu$, we find very consistent values with and without the \textit{Herschel} data included, indicative that this quality is robustly derived even without the \textit{Herschel} data. On the contrary, as expected, the dust temperature is fairly unconstrained without the Herschel coverage, with typical errors $\sim$4x larger when the \textit{Herschel} data are not included. Similarly, M$_{dust}$ is significantly less constrained without the \textit{Herschel} data.

We note two important aspects. First, this sample of 49 sources is by construction biased, including the brightest \textit{Herschel} sources. Therefore, these are arguably the sources for which including the Herschel photometry in the SED modeling would result in the largest effect compared to not including it. For fainter infrared sources, we expect the differences with and without the Herschel photometry to be smaller. Second, and most importantly, we stress that all our sources include \textit{Herschel} photometry, either as actual detections (of all or of a subset of the \textit{Herschel} bands), or upper limits. In particular, the inclusion of \textit{Herschel} upper limits in the SED modeling contributes enormously at constraining properties like SFR, A$_V$, L$_{dust}$, and f$_\mu$.

Therefore, we are confident that the MAGPHYS-derived quantities such as M$_*$, SFR, A$_V$, L$_{dust}$, and f$_\mu$ are robustly derived by our analysis. On the other hand, T$_{dust}$ and M$_{dust}$ are not as robust.

\begin{figure*}[h!]
\includegraphics[angle=-90, width=\columnwidth]{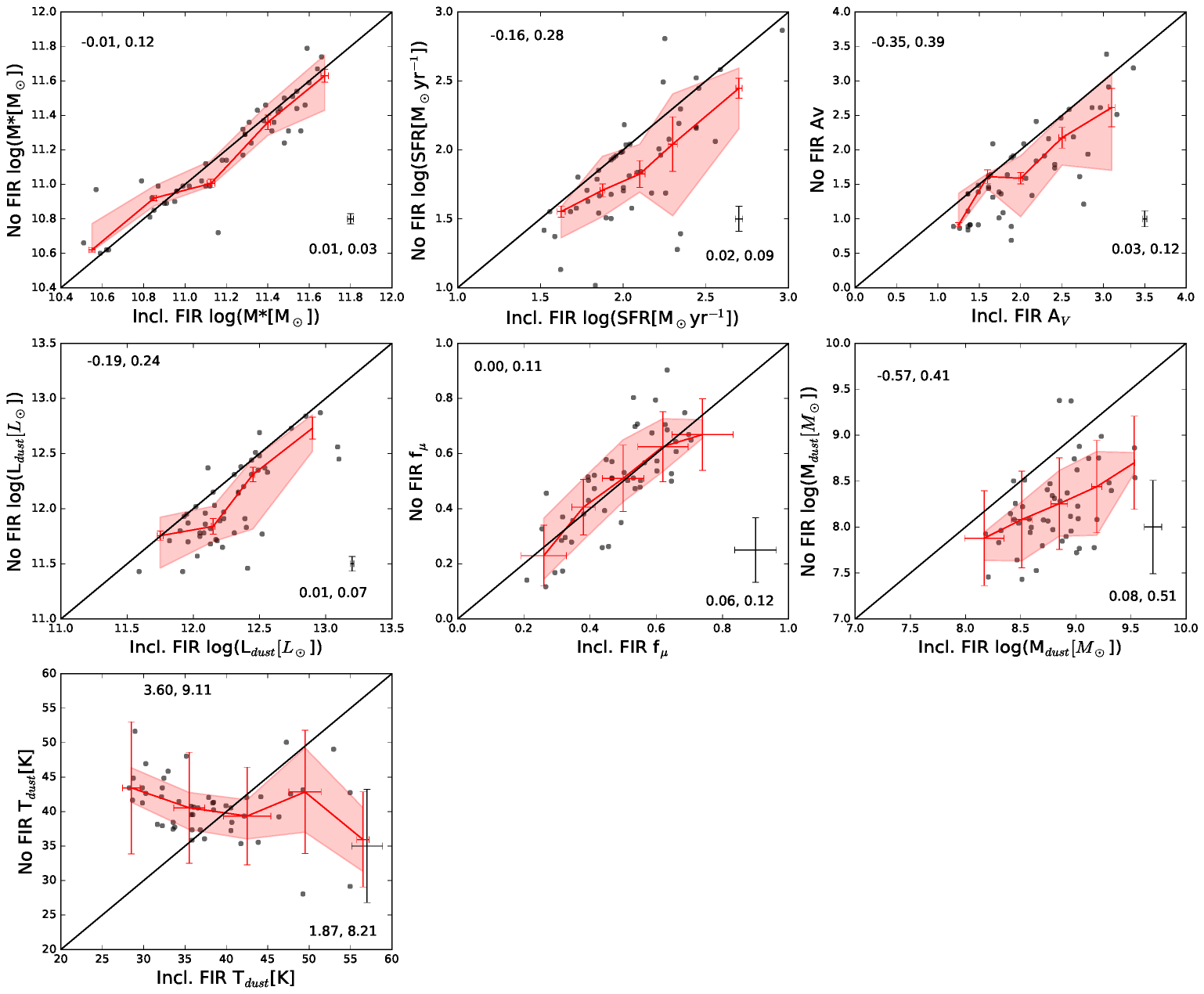}
\caption[width=\textwidth]{Comparison of MAGPHYS outputs obtained with and without including \textit{Herschel} photometry for the sources in our massive and dusty sample which are detected in all five \textit{Herschel} bands. The x-axis shows values obtained from including all available photometry in the modeling, the y-axis shows results from modeling of the the UV-MIPS SED. The red curve shows the running median with the shaded region indicating the range of the 16$^{th}$ to 84$^{th}$ percentiles. Red errorbars indicate the mean error for the objects in each bin. The mean errors for the full sample are shown in black and labeled by their magnitude in the given units. The median offset and scatter, respectively, for each comparison are shown at the top of their respective panels.  \label{A1}}
\end{figure*}

\section{B: Stellar Ages}

For completeness, here we present the MAGPHYS-derived mass-weighted stellar age. Figure 17 shows the distribution of the stellar age for the same samples plotted in Figure 1. Colors are as in Figure 1, with the final sample of massive and dusty galaxies plotted in red. Figure 18 shows the comparison of the stellar age with the specific star-formation rate (sSFR), f$_\mu$, the metallicity, and A$_V$ for the sample of massive and dusty galaxies, color-coded as a function of redshift. As expected, the stellar age anti-correlates with the sSFR, with older stellar ages for galaxies with smaller sSFR. Figure 18 all shows that the stellar age correlates with f$_\mu$, with older galaxies having higher values of f$_\mu$, consistent with older stellar populations having more of their dust emission originating from the diffuse ISM rather than the birth cloud. No correlation is instead found between the stellar age and the metallicity or A$_V$.

\begin{figure*}[h!]
\epsscale{0.8}
\plotone{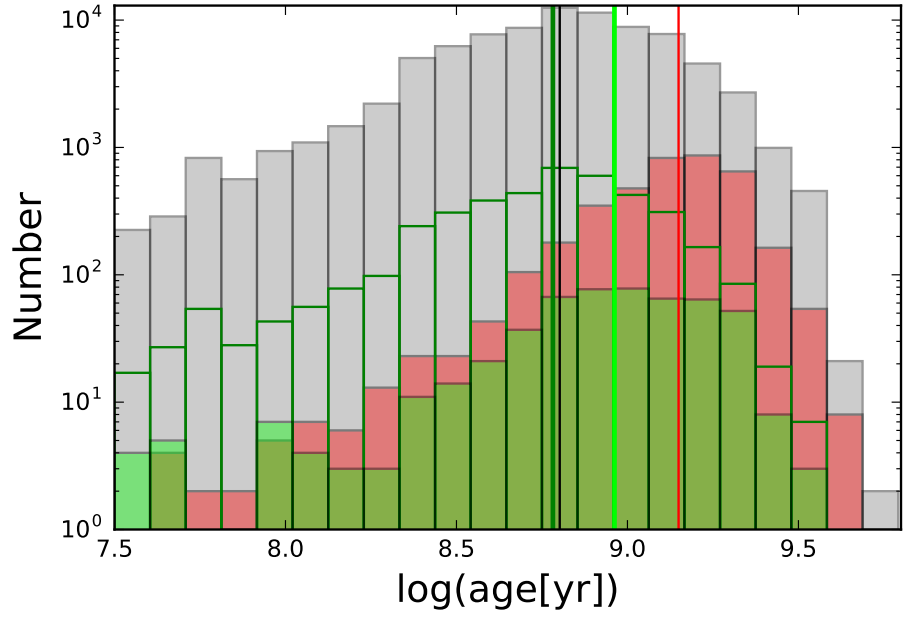}
\caption{Distribution of mass-weighted stellar ages derived by MAGPHYS for the same samples as shown in Figure 1.  \label{A1}}
\end{figure*}

\begin{figure*}[h!]
\plotone{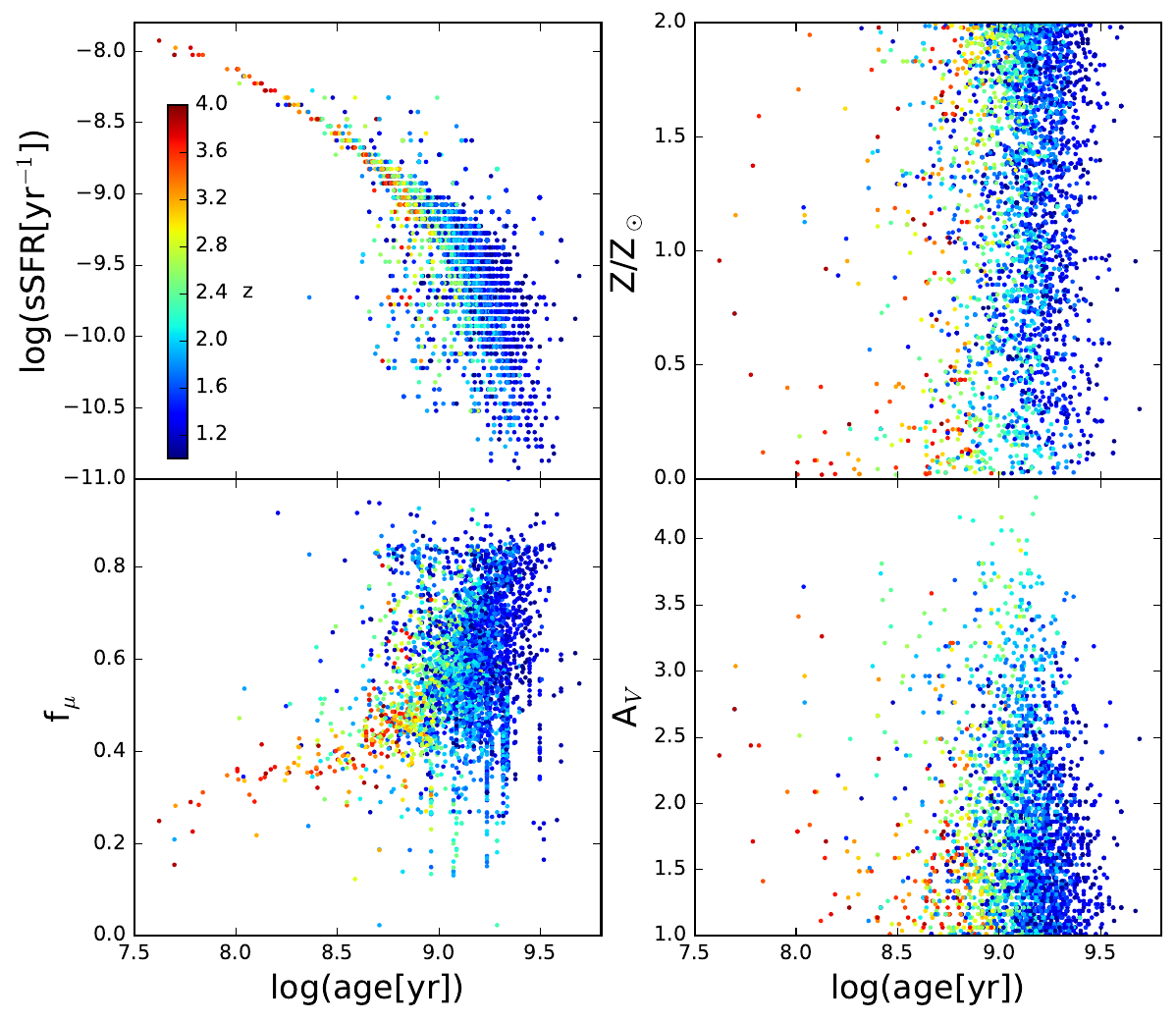}
\caption[width=\columnwidth]{Dependence of mass-weighted stellar age on sSFR, f$_\mu$, metallicity, and A$_V$ for the sample of massive and dusty galaxies. Coloring indicates redshift. \label{A1}}
\end{figure*}

\section{C: Alternate Classification of Star-forming and Quiescent Galaxies}

In order to investigate the dependence of the median SED shapes for star-forming and quiescent galaxies on our UVJ definition, we repeat the analysis for Figure 7 using a cut at $sSFR < 10^{-10}$yr$^{-1}$ to define our quiescent subsample. Figure 18 shows the results of this analysis. For comparison, the median SEDs using the UVJ color cut for star-forming and quiescent galaxies are shown in cyan and yellow respectively. The median SEDs for star-forming galaxies are rather robust against the selection criterion. For quiescent galaxies the median SEDs are also similar, as Figure 19 shows the SEDs for UVJ-selected sources fall within the shaded region showing the distribution of sSFR-selected SEDs, albeit with some differences. In the $2<z<2.5$ bin, the sSFR selection results in weaker UV emission and a shift to longer wavelengths of the FIR peak. At $2.5<z<3$, the sSFR selection results in this same shift of the FIR peak. At $3<z<4$ the alternate selection results in weaker UV emission and redder UV colors, although this bin now only includes three sources. Even with these differences, the qualitative results presented in the paper remain unchanged, so the comparison of median star-forming and quiescent SEDs for massive and dusty galaxies does not depend on the method used to select them.

\begin{figure*}[h!]
\includegraphics[width=\textwidth]{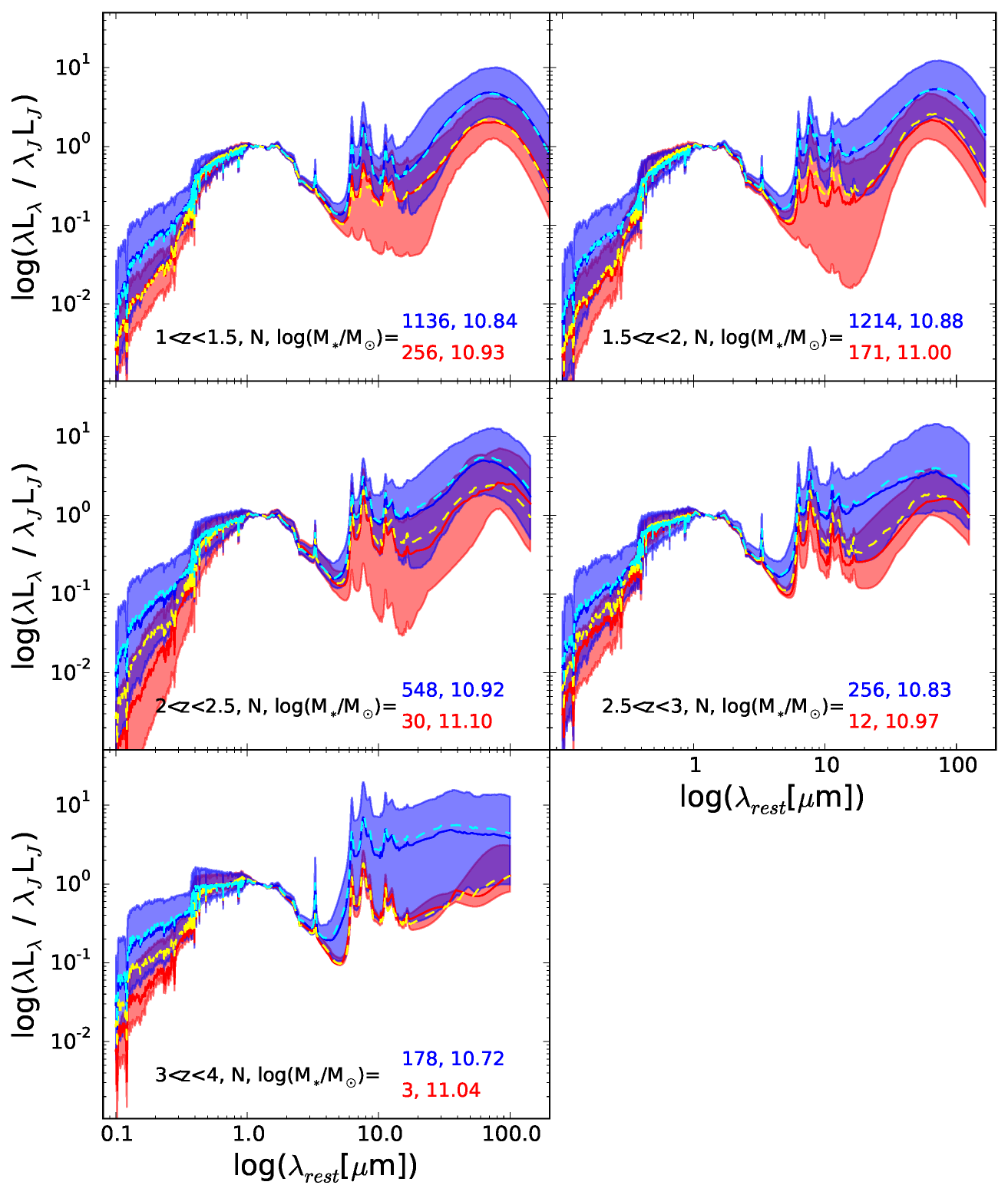}
\caption[width=\textwidth]{Median model SEDs for the entire sample with each panel corresponding to a different bin in redshift as labeled. Star-forming galaxies are shown in blue and quiescent galaxies in red defined by $sSFR < 10^{-10}$yr$^{-1}$, with the corresponding number and median stellar mass in each bin indicated. Shaded regions show the $15^{th}$ and $85^{th}$ percentiles of the distributions. Fluxes are normalized to the rest-frame J-band. For comparison, median SEDs for star-forming and quiescent galaxies selected by UVJ colors are shown in cyan and yellow respectively. \label{A1}}
\end{figure*}

\clearpage

\end{document}